\documentclass[preprint, 3p,times,twocolumn]{elsarticle}
\usepackage{color}

\usepackage{amssymb}
\usepackage[nodots]{numcompress}

\usepackage{lineno}

\journal{Computers \& Operations Research}

\usepackage[T1]{fontenc}
\usepackage[latin1]{inputenc}
\usepackage{lmodern}

\usepackage[ruled, linesnumbered, shortend]{algorithm2e} %linesnumberedhidden,
\SetAlCapHSkip{0em}

\usepackage{amsmath}
\usepackage{amssymb} %u.a. für Symbole von Zahlenmengen wie Natürliche Zahlen, Reele Zahlen
\usepackage{bm} %Vektoren fett darstellen

\usepackage{booktabs} % Anleitung: 'Publication quality tables', Simon Fear
\usepackage{array}

\usepackage{graphicx}
\usepackage{subfig}

\usepackage{multicol}
\newcolumntype{C}[1]{>{\centering}p{#1}}
\newcolumntype{R}[1]{>{\raggedleft}p{#1}}

\begin{document}

\begin{frontmatter}

%% Title, authors and addresses

%% use the tnoteref command within \title for footnotes;
%% use the tnotetext command for the associated footnote;
%% use the fnref command within \author or \address for footnotes;
%% use the fntext command for the associated footnote;
%% use the corref command within \author for corresponding author footnotes;
%% use the cortext command for the associated footnote;
%% use the ead command for the email address,
%% and the form \ead[url] for the home page:
%%
%% \title{Title\tnoteref{label1}}
%% \tnotetext[label1]{}
%% \author{Name\corref{cor1}\fnref{label2}}
%% \ead{tobias.buer@uni-bremen.de}
%% \ead[url]{home page}
%% \fntext[label2]{}
%% \cortext[cor1]{}
%% \address{Address\fnref{label3}}
%% \fntext[label3]{}

\title{A Pareto-metaheuristic for a bi-objective winner determination problem in a combinatorial reverse auction}

%% use optional labels to link authors explicitly to addresses:
%% \author[label1,label2]{<author name>}
%% \address[label1]{<address>}
%% \address[label2]{<address>}

\author{Tobias Buer\corref{cor1}}
\ead{tobias.buer@uni-bremen.de}
\cortext[cor1]{Corresponding author}

\author{Herbert Kopfer}
\ead{kopfer@uni-bremen.de}

\address{Chair of Logistics, University of Bremen, P.O. Box 330440, 28334 Bremen, Germany}

\begin{abstract}
The bi-objective winner determination problem (2WDP-SC) of a combinatorial procurement auction for transport contracts is characterized by a set $B$ of bundle bids, with each bundle bid~$b \in B$ consisting of a bidding carrier~$c_b$, a bid price~$p_b$, and a set~$\tau_b$ of transport contracts which is a subset of the set $T$ of tendered transport contracts.
Additionally, the transport quality $q_{t,c_b}$ is given which is expected to be realized when a transport contract $t$ is executed by a carrier~$c_b$.
The task of the auctioneer is to find a set~$X$ of winning bids ($X \subseteq B$), such that each transport contract is part of at least one winning bid, the total procurement costs are minimized, and the total transport quality is maximized.
This article presents a metaheuristic approach for the 2WDP-SC which integrates the greedy randomized adaptive search procedure with a two-stage candidate component selection procedure, large neighborhood search, and self-adaptive parameter setting in order to find a competitive set of non-dominated solutions.
The heuristic outperforms all existing approaches.
For seven small benchmark instances, the heuristic is the sole approach that finds all Pareto-optimal solutions.
For 28 out of 30 large instances, none of the existing approaches is able to compute a solution that dominates a solution found by the proposed heuristic.
\end{abstract}

\begin{keyword}
Pareto optimization \sep multi-criteria winner determination \sep combinatorial auction \sep GRASP \sep ALNS
%% keywords here, in the form: keyword \sep keyword

%% MSC codes here, in the form: \MSC code \sep code
%% or \MSC[2008] code \sep code (2000 is the default)

\end{keyword}

\end{frontmatter}
%%
%% Start line numbering here if you want
%%
%\linenumbers

\section{Introduction and literature review}
\label{sec:introduction}
Combinatorial auctions are applied when bidders are interested in multiple heterogenous items and when the bidders valuations of these items are non-additive.
This is for example the case with the procurement of transport services which often are highly interdependent.
We focus on these kinds of items in the following.
In a combinatorial transport auction, a shipper wants to procure transport services from many freight carriers. Items of a transport auction are denoted as transport contracts.
Such a contract is a framework agreement with a duration of about one to three years, that defines an origin location and a destination location between which a certain volume of goods has to be regularly carried (usually on the road)  while a specified service level has to be satisfied.

Combinatorial transport auctions allow freight carriers (bidders) to submit bundle bids.
A \emph{bundle bid} is an all-or-nothing bid on any subset of the set of tendered transport contracts.
In particular, a freight carrier can bid on combinations of transport contracts that exhibit strong synergies  (\cite{Kopfer_1999}, \cite{Pankratz_2000}, \cite{Sheffi_2004}). With this, the shipper strives to reduce his or her total transport costs.

Real-world applications of combinatorial auctions for the procurement of transport service are described by \citet{Ledyard_2002}, \citet{Elmaghraby_2004}, for example. \citet{Caplice_2006, Caplice_2003} discuss real-world issues of combinatorial transport auctions and report, among other things,  that practical transport auctions studied handle an average annual procurement volume of 150 million US-dollar. The whole auction process is complex and can last a few months \cite{Caplice_2006}.

After bidding is completed, the shipper (auctioneer) has to decide which of the received bundle bids should be accepted as winning bids. This problem is known as the winner determination problem which is usually modeled as a combinatorial optimization problem (for a review see \cite{Abrache_2007}). For combinatorial auctions which are used for selling items, the set packing problem is used to maximize the total revenue (compare \cite{Goossens_2009, Yang_2009}, for a review see \cite{Vries_2003}). Conversely, the winner determination problem of combinatorial procurement auctions like transport auctions are often modeled based on the set covering problem or the set partitioning problem and the total procurement costs are minimized.

In practice, shippers usually also want to ensure or improve service quality of the procured transport contracts ('transport quality') and therefore do not exploit their full potential for cost savings \cite{Sheffi_2004}. Models of winner determination problems of combinatorial auctions that try to integrate quality aspects in the decision making process are described in \cite{Caplice_2006}, \cite{Caplice_2003}, \cite{Catalan_2009}, \cite{Chen_2009}. Primarily, these approaches try to integrate quality aspects as some kind of side constraint or they use penalty costs to disadvantage low quality carriers or bundle bids, respectively. However, this  requires preference information of the shipper with respect to the desired trade-off between transport costs and transport quality. As \citet{Caplice_2006} state, identifying the desired trade-off is one of the most challenging tasks in the procurement of transport contracts for a shipper.
Therefore, \citet{Buer_2010a} introduced an additional, second objective function for maximizing the transport quality within a winner determination problem. The resulting bi-objective model, denoted as 2WDP-SC, seems helpful, if the desired trade-off between transport costs and transport quality is a priori unknown to the shipper.

To solve multiple criteria optimization problems (combinatorial and continuous) without an a priori known preference information there are several approaches like MOGLS \cite{Ishibuchi_1998, Jaszkiewicz_2002}, NSGA-II \cite{Deb_2002}, SPEA2 \cite{Zitzler_2002}, PMA \cite{Jaszkiewicz_2004}, MOEA/D \cite{Zhang_2007}, or EMOSA \cite{Li_2011}. These methods focus on the multiple criteria aspect of the optimization problems and have been successfully applied to continuous optimization problems, but also to combinatorial problems like Knapsack or Traveling Salesman.
The goal of this paper is \emph{not} to provide a problem-independent multiple criteria solver like the ones mentioned. Instead, our heuristic focuses on the solution of 2WDP-SC with the hope to trade off a narrower applicability against improved solution quality.
Nevertheless, the algorithmic ideas presented might prove helpful for the solution of other set covering based multiple criteria problems.

This paper continues the work of \cite{Buer_2010a, Buer_2010b, Buer_2011} and proposes a new heuristic solution approach for the 2WDP-SC. Our previous approaches include epsilon branch-and-bound to solve the 2WDP-SC to optimality, a genetic algorithm, and matheuristics based on GRASP and path relinking combined with the epsilon branch-and-bound procedure.
The present paper introduces a heuristic called Pareto neighborhoodsearch (PNS) that is based on a multi start procedure followed by large neighborhood search. Differences from our previous work are:

\begin{enumerate}
  \item The multi start construction process uses a two-stage candidate component selection procedure in order to generated a diverse set of non-dominated solutions.
  \item During large neighborhood search we now use multiple destroy rates and two repair heuristics that privilege individual objective criteria on a rotating basis. The use of these operators is controlled self-adaptively and depends on the invested effort to find a new non-dominated solution starting from a specific region in the objective space.

  \item PNS is less complicated compared to \cite{Buer_2010b, Buer_2011}, there is no post construction optimizer, no path relinking, and no branch-and-bound.
  \item To evaluate PNS, we additionally present new computational results for three existing approaches. We computed 37 instances again on the same computer, used an extended set of quality indicators, and performed a runtime analysis including empirical runtime distributions.
\end{enumerate}

The heuristic PNS clearly outperforms all previously existing approaches, i.e., for 35 out of 37 instances the competing heuristics cannot even generate a sole solution that dominates any of the solutions generated by PNS. PNS is the second fastest approach tested. Furthermore, we demonstrate that the proposed concepts of the construction phase as well as in the neighborhood search phase contribute to find superior solutions.
Some of the presented results are also part of the German language Ph.D thesis \cite{Buer_2012b}.
The article is organized as follows. Section \ref{sec:problem} introduces the studied bi-objective winner determination problem. To solve it, we present a new Pareto metaheuristic called PNS (Section \ref{sec:algo}). The performance of PNS is evaluated by means of a benchmark study (Section \ref{sec:study}) whose results are discussed in Section \ref{sec:results}. Section \ref{sec:conclusion} summarizes the findings.

\section{The bi-objective winner determination problem}
\label{sec:problem}
The bi-objective winner determination problem of a combinatorial transport procurement auction based on a set covering formulation (2WDP-SC) has been introduced by \citet{Buer_2010a}.
We are given a set $T$ of transport contracts offered by a single shipper (decision maker) and a set $B$ of bundle bids which have been submitted by a set $C$ of carriers. A bundle bid $b \in B$ is composed of a carrier $c_b \in C$, a bid price $p_b \in \mathbb{R^{+}}$, and a subset $\tau_b$ of the offered transport contracts $T$. With the bundle bid $b$, the carrier $c_b \in C$ expresses the intention to execute the set of transport contracts $\tau_b \subseteq T$, if he gets paid the price $p_b$ by the shipper. Let $a_{tb} = 1$ if $t \in \tau_b$ and $a_{tb} = 0$ otherwise ($\forall t \in T, \forall b \in B$). If $a_{tb}=1$, we say $b$ \emph{covers} $t$.
Furthermore, we are given parameters $q_{t,c_b}\in \mathbb{N}$ ($\forall t \in T, c \in C$) which indicate the achieved transport quality if transport contract $t$ is executed by carrier $c$ who submitted bundle bid $b \in B$. The shipper prefers higher values of $q_{t,c_b}$.

The optimization task of the shipper is to determine a set of winning bids~$X$ ($X \subseteq B$).
The binary decision variable $x_b$ indicates, whether bundle bid $b \in B$ is accepted as winning bid ($x_b=1 \Leftrightarrow b \in X$) or not.
The 2WDP-SC asks for the set of winning bids $X$ that covers all transport contracts $T$ and at the same time strives to do both, to minimize the total procurement costs and to maximize the total transport quality. The 2WDP-SC is defined by the expressions (\ref{eq:ip_f1}) -- (\ref{eq:ip_binary}).

\begin{align}
\label{eq:ip_f1}
\min f_1(X)  = &\sum_{b \in B} p_b \cdot x_b,  \\
\label{eq:ip_f2}
\min {f}_2(X) = &(-1)\sum_{t \in T} \max_{b \in B} \{q_{t,c_b} \cdot a_{tb} \cdot x_b \},  \\
\label{eq:ip_cover}
\text{s.\,t.}\quad\quad\;\;\;\;\;    &\sum_{b \in B} a_{tb} \cdot x_b \geq 1, \;\; \quad\quad\quad \forall t \in T, \\
\label{eq:ip_binary}
&x_b \in \{0,1\},   \quad\quad\quad\quad\quad\;\; \forall b \in B.
\end{align}
Objective function $f_1$ (\ref{eq:ip_f1}) minimizes the total procurement costs of the shipper. That is, the sum of the prices of the winning bids. Objective function $f_2$ (\ref{eq:ip_f2}) maximizes the total transport quality of the procured transport contracts. For ease of notation used later, we minimize the negative total transport quality to obtain a pure minimization problem. Constraint set (\ref{eq:ip_cover}) guarantees, that each transport contract is covered by \emph{at least one} winning bid. Finally, expression (\ref{eq:ip_binary}) ensures, that each bundle bid is an all-or-nothing bid, that is, partial acceptance of a bundle bid is prohibited.

The formulation of the objective function $f_2$ is influenced by the set covering inequality (\ref{eq:ip_cover}). Because of (\ref{eq:ip_cover}), a transport contract $t$ may be covered by multiple winning bids although it must be executed only once. Therefore, the maximum function in $f_2$ makes sure, that for each transport contract $t$ only the highest transport quality value $q_{t,c_b}$ for the given set of winning bids is summed up once.
Alternatively, the shipper could forbid the coverage of a contract by multiple winning bids, that is, instead of $\geq$ the $=$-operator could be used in (\ref{eq:ip_cover}). However, this additional restriction \emph{cannot} lead to solutions with lower total procurement costs. In contrast, finding a feasible solution is more complicated because a set partitioning problem has to be solved instead of a set covering problem. The total procurement costs may therefore even increase. Using the $\geq$-operator in (\ref{eq:ip_cover}) requires the free disposal assumption \cite{Sandholm_2002} to hold, that is, a carrier will \emph{not} charge additional costs for executing less contracts than offered in a bundle bid. This appears plausible in the scenario at hand. For a more detailed discussion of using a set covering or set partitioning formulation in this context please see \cite[p. 195f]{Buer_2010b}.

The expressions (\ref{eq:ip_f1}), (\ref{eq:ip_cover}), and (\ref{eq:ip_binary}) define the well-known NP-hard set covering problem \cite{Karp_1972}. If a single objective decision problem with $f_k, k=1$ is NP-complete, then the corresponding multi objective decision variant with $f_k, k>1$ is also NP-complete \cite{Serafini_1986}. Therefore, the 2WDP-SC is NP-hard.

Finally, we introduce the notation of solution dominance.
Let $k$ be the number of objective functions of a minimization problem and let $X^1, X^2$ be two feasible solutions. $X^1$ \emph{weakly dominates} $X^2$, written $X^1 \preceq X^2$, if $f_i(X^1) \leq f_i(X^2), i = 1, \ldots, k$. $X^1$ \emph{dominates} $X^2$, written $X^1 \prec X^2$, if $f_i(X^1) \leq f_i(X^2), i = 1, \ldots, k$ and $f_i(X^1) < f_i(X^2)$ holds at least for one $k$.  An \emph{approximation set} $A$ is a set of feasible solutions which do not $\prec$-dominate each other. The approximation set which contains those feasible solutions which are not weakly dominated by any other feasible solution is called Pareto-optimal set.

\section{A Pareto metaheuristic based on GRASP and adaptive LNS}
\label{sec:algo}

We denote our solution approach as Pareto neighborhood search (PNS). An overview is given by Alg. \ref{alg:paretometaheuristic}. The approach consists of a construction phase inspired by ideas of the greedy randomized adaptive search procedure (GRASP, \cite{Feo_1995}) and an improvement phase based on concepts known from adaptive large neighborhood search (ALNS, \cite{Ropke_2006}).
Some of the basic concepts adopted from GRASP and ALNS are adjusted and extended to cope with multiple objective criteria. Acceptance of new solutions during the search is always based on the Pareto dominance principle.

PNS requires as input a set of bundle bids $B$, which defines the problem instance, and four algorithmic parameters $l^{max}$, $time\_limit$, $s$, and $\bm d$. The parameter $time\_limit$ is used as termination criterion of PNS and controls the permitted runtime. The remaining parameters are explained below together with the details of the construction phase (DRC) and the improvement phase (PLNS).

\begin{algorithm}[!htb]
\caption{PNS -- Pareto neighb. search}
\label{alg:paretometaheuristic}

\SetAlgoLined
\DontPrintSemicolon

\SetKwFunction{constructSolution}{DRC}
\SetKwFunction{localSearch}{PLNS}

\KwIn{$B$, $s$, $l^{max}$, $\bm d$}
\KwOut{approximation set $A$}
\BlankLine
$A \leftarrow \constructSolution(B, s, l^{max}) $ \tcp*{cf. Alg. \ref{alg:construct}}
$A \leftarrow \localSearch(B, A, \bm d)$ \tcp*{cf. Alg. \ref{alg:plns}}
\Return $A$ \;
\end{algorithm}

\subsection{Construction Phase (DRC)} \label{sec:alg_construction}

The construction phase of PNS is denoted as \emph{dominance-based randomized construction} (DRC, cf. Alg. \ref{alg:construct}).
The goal of DRC is to generate a diverse set of non-dominated solutions. DRC is inspired by the construction phase of the multi-start metaheuristic GRASP. In general, each component that can be integrated in the incomplete solution being built is evaluated by a greedy function. A subset of the best components is formed by the so called restricted candidate list. From this restricted candidate list a component is randomly drawn and inserted in the solution being built. These steps are repeated and the process terminates after the built solution is feasible. Usually, immediately after constructing a feasible solution it is improved via local search. However, we do not follow this pattern. From experiments with the approach of \cite{Buer_2010b} we learned that our multi criteria local search is more effective if it is applied to a set of non-dominated solutions instead of a single solution.
Further modifications arise through the multi criteria nature of the problem at hand and affect the formation of the candidate list as well as the component selection process.
Despite being originally developed for single objective optimization, GRASP seems well-suited for the task at hand. To generate a set of non-dominated solutions, multiple solutions have to be constructed which is a major characteristic of GRASP. A constructive approach like GRASP is also favored by the problem at hand, because it is usually easy to find feasible solutions for set covering based problems.
For example, the randomized greedy multi start approach of \cite{Lan_2007} is  among the most competitive solution approaches for the single criteria set covering problem.

\subsubsection{Two-stage component selection procedure}

DRC is a multi start procedure that obtains an approximation set by iteratively constructing feasible solutions.
Solutions of combinatorial optimization problems are made up of components. The components of a solution for the 2WDP-SC are bundle bids $b \in B$.
DRC requires a greedy function to rate components (i.e., bids from $B$). Only this function is problem-specific and described in Sect.~\ref{sec:algo_greedy}, all the rest of the proposed procedures are problem-independent. Nevertheless, we use the concrete term bundle bid instead of the term component to introduce the procedure.

The procedure DRC constructs a feasible solution in each iteration of the repeat-until loop (cf. Alg. \ref{alg:construct}, lines \ref{algl:drc_repeat} -- \ref{algl:drc_until}).
At first, some variables are initialized or updated: $k$ counting the number of constructed solutions (line 5); the solution under construction $X$ (line 6); and a temporary working copy $B'$ of the set of bundle bids $B$ (i.e., all potential components of a solution).
In lines  8--12 a subset of bundle bids which forms a feasible solution is selected.

\begin{algorithm}[!htb]
\caption{DRC -- construction phase}
\label{alg:construct}

\SetKwFunction{cl}{genCandList}
\SetKwFunction{selCand}{selCandSector}

\SetAlgoLined
\DontPrintSemicolon

\KwIn{$B$, $s$, $l^{max}$}
\KwOut{approximation set $A$}
\BlankLine
$A \leftarrow \emptyset$ \;
$k \leftarrow 0$ \;
$l \leftarrow 1$  \;
\Repeat{$l = l^{max}$}{ \nllabel{algl:drc_repeat}
%\tcp*[f]{restart loop}
%
%
%
$k \leftarrow k + 1$ \;
$X \leftarrow \emptyset$ \;
$B' \leftarrow B$ \;
\While{$X$ infeasible}{
    \tcp{first\,stage, cf.\,Alg.\,\ref{alg:candidateList}}
    $C \leftarrow \cl(B', X)$   \;
    \tcp{second\,stage,\,cf.\,Alg.\,\ref{alg:selectCandidate}}
	$b' \leftarrow \selCand(C, k, s)$    %\;
	$X \leftarrow X \cup \{b'\}$  \;

}

\uIf{$(\not \exists X' \in A | X' \prec X)$}{ \nllabel{algl:drc_dom}
$A \leftarrow A \uplus \{ X \}$ \;
$l \leftarrow 1$ \;
}
\Else{
$l \leftarrow l + 1$ \;
}
}
\nllabel{algl:drc_until}
\Return $A$ \;
\end{algorithm}

In order to select appropriate bundle-bids (i.e., components), we propose a \emph{two-stage component selection procedure} which takes multiple criteria into account and distinguishes DRC from standard GRASP as well as from the previous multi-criteria approach for the 2WDP-SC of \cite{Buer_2010b}.

The \emph{first stage} is realized by the procedure \emph{genCandList} (cf. Alg. \ref{alg:construct}, line 9 and Alg. \ref{alg:candidateList}) which computes a candidate list $C, C \subseteq B'\setminus X$, of potential bundle bids that can be added to the constructed solution $X$. For this purpose, each candidate bid $b \in B'\setminus X$ is rated by a vector-valued greedy function $\bm g(b,X)$.
This function calculates a greedy rating vector for $b$.
Each component of the rating vector corresponds to an objective function of the optimization problem.
Let us agree that lower values of the components of $\bm g(b,X)$ are superior and a value of $+\infty$ signals no contribution of $b$ to improve $X$ at all. For each optimization problem, a specific greedy function $\bm g(b,X)$ has to be adapted; for the 2WDP-SC at hand, the chosen function is explained in Sect.~\ref{sec:algo_greedy}.

If a bid makes no contribution at all, it is not necessary to rate it again in further iterations (cf. Alg. \ref{alg:candidateList}, line 3). On the other hand, if a bid makes a contribution we add it to the candidate list $C$ using the operator $\uplus$ (cf. Alg.~\ref{alg:candidateList}, line 5). The operator $\uplus$ symbolizes, that a bid $b$ is only added to $C$ if $b$ is not dominated by any of the bundle bids in $C$.
After adding $b$ to $C$, those bundle bids in $C$ which are dominated by $b$ are removed from $C$.

\begin{figure}
  \includegraphics[width=0.45\textwidth]{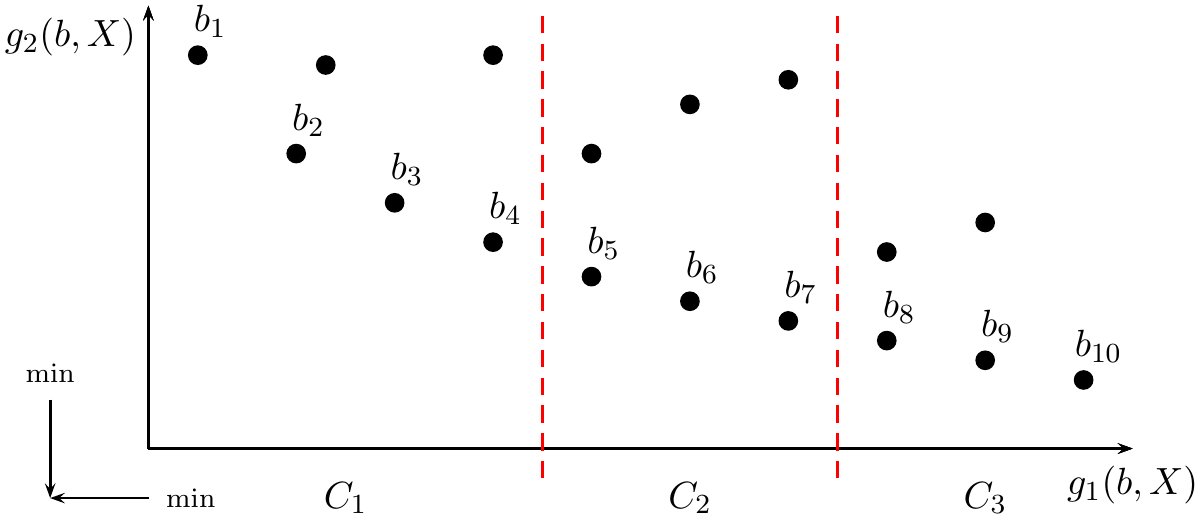}
  \caption{Organization of candidate list $C$, $C = \{b_1, \ldots, b_{10}\}$, and $r=3$.}
  \label{fig:candidatelist}
\end{figure}

Fig. \ref{fig:candidatelist} shows an example. The 17 points represent the greedy rating vectors of 17 bundle bids which are rated by $\bm g(b,X) = (g_1(b,X), g_2(b,X))$ (see Sect. \ref{sec:algo_greedy}).
Each unnumbered point is dominated by at least one numbered point $b_1, \ldots, b_{10}$. The numbered points (bundle bids) are non-dominated with respect to each other. Therefore, the bids $b_1, \ldots, b_{10}$ constitute the candidate list $C$ defined in the first stage of the component selection procedure.

\begin{algorithm}[!htb]
\caption{genCandList -- first stage}
\label{alg:candidateList}

\SetAlgoLined
\DontPrintSemicolon

\KwIn{$B', B' \subseteq B$, $X$}
\KwOut{candidate list $C, C \subseteq B'$}
\BlankLine
\ForEach{$b \in B' \setminus X$}{
	\uIf
	{${\bm g(b,X)} = {\bm \infty}$}
	{$B' \leftarrow B' \setminus \{b\}$ } %\tcp*[r]{$b$ nicht mehr bewerten} }
	\Else{
         $C \leftarrow C \uplus \{b\}$ \;
	}
}
\Return $C$ \;
\end{algorithm}

The \emph{second stage} is realized by the procedure \emph{selCandSector} (cf. Alg. \ref{alg:construct}, line 10 and Alg.~\ref{alg:selectCandidate}) which divides the candidate list into so called sectors.
Required input of the procedure are the candidate list $C$, the number of constructed solutions $k$, and the external parameter $s \in \mathbb{N}$.

First, the bundle bids of the given candidate list $C$ are partitioned into $s$ subsets $C_1, \ldots, C_s$ which are denoted as \emph{sectors}.
All sectors should contain the same number of bundle bids. If an equal division of bids to sectors is not possible, then left over bids are assigned to the first sector $C_1$. The partitioning of bundle bids into the $s$ sectors happens implicitly in Alg.~\ref{alg:selectCandidate}, lines~1~--~5. The bids of $C$ are sorted and numerated in ascending order of $g_1(.,X)$, i.e. $g_1(b_1,X) \leq g_1(b_2,X) \leq \ldots \leq g_1(b_{|C|})$. The cardinality $m_1 = |C_1|$ as well as the cardinalities $m_2 = |C_2|, \ldots, m_r=|C_s|$ are calculated assuming $m_1 \geq m_2 = \ldots = m_r$. In the example of Fig.~\ref{fig:candidatelist}, the values are $m_1 = 4, m_2 = m_3 = 3$.

\begin{algorithm}[!htb]
\caption{selCandSector -- second stage}
\label{alg:selectCandidate}

\SetAlgoLined
\DontPrintSemicolon

\KwIn{$C$, $k$, $s$}
\KwOut{a bundle bid~$b, b \in C$}
\BlankLine

sort all $b \in C$ in ascending order of $g_1(b,X)$ \;

$n \leftarrow |C|$ \;
\lIf{$s>n$}{$s \leftarrow n$}

$m_j \leftarrow \left\lfloor n/s \right\rfloor$
\tcp*[r]{$|C_j|, 2 \leq j \leq s$}

$m_1 \leftarrow n - m_j \cdot (s-1)$
\tcp*[r]{$|C_1|$}

\ShowLn\label{algl:4-selectCandidate-sektor}
$i \leftarrow k$ mod $s$ \tcp*[r]{choose sector}

\uIf{$i = 1$}{
	$C_i \leftarrow C[1, m_1]$
}
\Else{
	$C_i \leftarrow C[m_1 + m_j \cdot (i-2) +1, m_1 + m_j \cdot (i-1)]$
}

select a bid $b \in C_i$ with probability $1/|C_i|$ \;
\Return $b$ \;
\end{algorithm}

Second (cf. Alg.~\ref{alg:selectCandidate}, lines 6 -- 11), the sector $C_i$ is selected by $i = k \mod s$; the number of up to now constructed solutions $k$ is modulo divided by the number of sectors $s$. Then the bids in $C_i$ are determined. The notation $C[1,m_1]$ refers to the set of elements of the sorted list $C$ from position $1$ to position $m_1$ (inclusive). In the example of Fig.~\ref{fig:candidatelist}, in case of $i=2$, we get $C_2 = C[5,7] = \{b_5, b_6, b_7\}$. Finally, a bundle bid $b$  is randomly drawn from the sector $C_i$ with probability $1/|C_i|$ (cf. Alg.~\ref{alg:selectCandidate}, line 12).

In other words, DRC constructs the $k$th solution with bundle bids from the $(k \mod s)$th sector and so forth. With $s=3$, all bundle bids for the first solution ($k=1$) are drawn from $C_1$. The second solution $(k=2)$ is constructed with bundle bids from $C_2$ etc. In doing so, the assignment of bundle bids to sectors is always updated after adding a bundle bid to the solution under construction.

The idea of segmenting the bundle bids into sectors is to smoothly guide the search process into certain directions of the multi criteria objective space. Thus, all variants of (non-dominated) compromise solutions are constructed and not only a variation of extreme solutions with respect to a single objective function (e.g., as in the approach of \cite{Buer_2010b}). This important effect is achieved without the need to configure weights for the objective functions, only the number of sectors $s$ has to be chosen which is, at least for the 2WDP-SC, a straight-forward task.
Without dividing bundle bids of the candidate list to several sectors (assume $C_i := C$) the consecutive decisions which bundle bid to add to the constructed solution might more easily contradict. For example, if in the first step a bid is selected from the candidate list which heavily favors $f_1$ and in a later decision a bid is selected from the updated candidate list which heavily favors $f_2$, then this might not lead to a good comprise solution but simply to a bad solution. Applying the sector based approach,  the decisions are stronger balanced with respect to the conflicting objectives without the need of weighting the objectives.

\subsubsection{Greedy rating functions}
\label{sec:algo_greedy}
The vector-valued greedy rating function $\bm g(b,X)$ has to be designed with respect to the problem at hand. Here, we use the greedy function $\bm g(b,X) = (P(b,X), Q(b,X))$ as introduced in \cite{Buer_2010b}.

The rating function $P(b,X)$ is defined according to (\ref{eq:greedyP}). It measures the ability of a bundle bid $b \not\in X$ to make $X$ a feasible solution and to improve $f_1(X)$.
Let $\tau(X)$ denote the set of contracts covered by $X$, i.e. $\tau(X) = \bigcup_{b \in X} \tau(b)$.
$P(b,X)$ calculates the average additional costs for those contracts in $b$ which are not yet covered by $X$ (cf. \citet{Chvatal_1979}).
If all transport contracts $\tau(b)$ of a bundle bid $b$ are already covered by $X$, then $b$ cannot contribute to reach feasibility of $X$ and therefore $P(b,X) = +\infty$.
Lower values of $P(b,X)$ are considered as better.

\begin{align}
\label{eq:greedyP}
P(b,X) &=
\begin{cases}
\frac{p_b}{\mid\tau(b) \setminus \tau(X)\mid} & \text{if } \tau(b) \setminus \tau(X) \neq \emptyset, \\
+\infty & \text{otherwise.}
\end{cases}
\end{align}

The rating function $Q(b,X)$  defined according to (\ref{eq:greedyQ}) measures the ability of a bundle bid $b \not\in X$ to improve $f_2(X)$.
By accepting an additional bundle bid $b$ as winning bid the transport quality $f_2$ may only increase, i.e., $\Delta f_2(X) = f_2(X \cup \{b\}) - f_2(X) \geq 0$.
In contrast to $f_1$, the value of $f_2$ cannot worsen by accepting an additional bid.
The increment in transport quality $\Delta f_2(X)$ is divided by the total number of contracts covered by each individual bid in $X \cup \{b\}$ , that is $\sum_{b' \in X \cup \{b\}} |\tau(b')|$. Hence, covering a contract by several bids is penalized.
Finally, this value is multiplied by $-1$, so that smaller values of $Q(b,X)$ represent better bids.

\begin{align}
\label{eq:greedyQ}
Q(b,X) &=
\begin{cases}
-\dfrac{\Delta f_2(X)}{\sum_{b' \in X \cup \{b\}} {\mid\tau(b')\mid}} & \text{if } \Delta f_2(X) > 0, \\
+\infty
 & \text{otherwise.}
\end{cases}
\end{align}

\subsubsection{Termination criterion}
The multi start construction procedure terminates, if $l^{max}$ solutions are constructed successional without finding a new non-dominated solution.
The checks are performed by Alg. \ref{alg:construct}, in Lines 3 and \ref{algl:drc_dom} -- \ref{algl:drc_until}.

\subsection{Improvement phase (PLNS)} \label{sec:alg_neigbhorhood}
In order to generate competitive solutions from the approximation set found during the construction phase, an improvement phase is applied.
The improvement phase (cf. Alg. \ref{alg:plns}) is inspired by large neighborhood search \cite{Shaw_1998} with adaptive parameters \cite{Ropke_2006}.
The adaptive large neighborhood search (ALNS) metaheuristic defines the neighborhood of a solution as the set of solutions that can be generated by applying a destroy operator to the solution and then a repair operator. Usually, the destroy operator works in a randomized fashion. Based on the destroyed solution the repair operator generates a new feasible solution. For this, greedy heuristics as well as exact branch-and-bound methods may be applied. The new solution may be accepted as a new candidate solution for continuing search even if it is inferior to the best known solution in order to escape local optima. ALNS makes use of several destroy and repair operators. The success of each operator is tracked and influences the probability that an operator is used again during the search.
ALNS usually terminates, after reaching a preset time limit.
Modifications of this pattern are mainly due to the multi criteria setting which in particular influences the success tracking and selection of operators.

\subsubsection{Outline and self-adaptive parameter setting}
The destroy and repair principle with self-adaptive parameters is designed as follows.
A solution $X \in A$ is chosen (cf. Alg.~\ref{alg:plns}, line 2). The destroy heuristic randomly removes some bundle bids and returns a (most likely) infeasible solution $X^d$ (cf. Alg.~\ref{alg:plns}, line 3). $X^d$ is repaired via a greedy heuristic and the resulting feasible solution is denoted as $X^r$ (cf. Alg.~\ref{alg:plns}, line 4). In line 5 of Alg.~\ref{alg:plns} the PLNS procedure checks whether $X^r$ is a new non-dominated solution.
If it is not, that is, the attempt to improve $X$ with the given parameters of the destroy and the repair heuristic failed, then either $\sigma_1(X)$ or $\sigma_2(X)$ is incremented (lines 9 -- 13).

The values $\sigma_1(X), \sigma_2(X) \in \mathbb{N}, \forall X \in A,$ control the self-adaptive parameter setting of the destroy heuristic as well as of the repair heuristic. The destroy heuristic adapts the destroy rate (that is, the probability to remove a given bundle bid from a solution) depending on a given destroy strategy as well as on $\sigma_1(X)$, and $\sigma_2(X)$ (cf. Alg.~\ref{alg:destroySol}, line 1). Furthermore, the repair heuristic uses $g_1$ as rating function if $\sigma_1(X) < \sigma_2(X)$ and $g_2$ otherwise (see Alg. \ref{alg:repairSol}, line 2). Thus, $\sigma_1(X)$ and $\sigma_2(X)$ are counting functions, that count the number of failed attempts to improve solution $X$ while using $g_1(.,.)$ and $g_2(.,.)$ as greedy repair criteria, respectively.

We note that the failed improvement attempts counters $\sigma_1(X)$ and $\sigma_2(X)$ do not depend on the approximation set but on individual non-dominated solution $X \in A$.
This seems reasonable, as the non-dominated solutions in $A$ often differ strongly in the decision space.
That is, the values of corresponding decision variables of non-dominated solutions differ frequently.
Even stronger structural differences on the decision space level may occur for solutions that lie in very different areas of the objective space but are nevertheless non-dominated. For example, our experience shows that a 2WDP-SC solution $X$ with a high $f_1(X)$ and a low $f_2(X)$ contains more and different bundle bids compared to another solution $X'$ with a low $f_1(X')$ and a high $f_2(X')$.
With the focus on individual solutions of $A$, the improvement phase is able to adapt the used parameters to structural differences of the non-dominated solutions.

\begin{algorithm}[!htb]
\caption{PLNS -- improvement phase}
\label{alg:plns}

\SetAlgoLined
\DontPrintSemicolon

\KwIn{approximation set $A$, $\bm d = (d_1, \ldots, d_n)$}
\BlankLine
\SetKwFunction{d}{destroySol}
\SetKwFunction{r}{repairSol}
\SetKwData{fP}{$\sigma_1(X)$}
\SetKwData{fPr}{$\sigma_1(X^r)$}
\SetKwData{fQ}{$\sigma_2(X)$}
\SetKwData{fQr}{$\sigma_2(X^r)$}

\Repeat{time limit reached}{

    pick a solution $X \in A$ \label{algl:plns-remove} \;

	$X^d \leftarrow $ \d{$X$, \fP, \fQ, $\bm d$} \;
	
	$X^r \leftarrow $ \r{$X^d$, \fP, \fQ} \;
	
	\uIf{$\nexists X^a \in A \mid X^a \preceq X^r$}{
        $A \leftarrow A \uplus \{X^r\}$ \label{algl:plns-insert} \; %\tcp*[r]{insert at back of $A$}
		$\fPr \leftarrow 0$ \;
		$\fQr \leftarrow 0$ \;
	}
	\uElseIf{$\fP < \fQ$}{
		$ \fP \leftarrow \fP + 1 $ \;
	}
	\Else{
		$ \fQ \leftarrow \fQ + 1 $ \;
	}
}

\Return $A$ \;
\end{algorithm}

\subsubsection{Destroy procedure with adaptive destroy rates}
The procedure \emph{destroySol} (cf. Alg. \ref{alg:destroySol}) randomly removes bundle bids from a given solution $X$  with a specified removal probability, denoted as destroy rate.
The destroy rate is the probability by which a bundle bid $b \in X$ is removed from $X$.
A vector of destroy rates is denoted as destroy strategy $\bm d \in [0,1]^n, n \geq 1$.
The destroy strategy $\bm d$ is a given parameter, however, which  destroy rate $d_i, 1 \leq i \leq n$ of the given strategy is applied at a given time is decided in a self-adaptive manner.

The procedure $rand(1,100)$ used in line 3 of Alg.~\ref{alg:destroySol} returns a random number between 1 and 100 (inclusively).

\begin{algorithm}[!htb]
\caption{destroySol}
\label{alg:destroySol}

\SetAlgoLined
\DontPrintSemicolon
\SetKwData{fP}{$\sigma_1$}
\SetKwData{fQ}{$\sigma_2$}
\SetKwData{rand}{rand}

\KwIn{$X$, \fP, \fQ, $\bm d = (d_1, \ldots, d_n)$ }
\BlankLine
$i \leftarrow \min(\fP, \fQ) \mod n$ \label{algl:destroySol-select} \;
\ForEach{$b \in X$}{
	\lIf{\rand(1,100) $\leq d_i$}{$X \leftarrow X \setminus \{b\}$}
}

\Return $X$   \;
\end{algorithm}

\subsubsection{Repair procedure}

The procedure \emph{repairSol} (cf. Alg. \ref{alg:repairSol}) uses a single-criterion greedy heuristic to repair an infeasible solution $X^d$. Further input is $\sigma_i \in \mathbb{N}, i = 1,2$ which denotes the number of \emph{failed attempts} to find a new non-dominated solution with respect to $A$ provided that the search started with $X$ and the operator $g_i$ was used.
Compare this with the input $\sigma_1(X)$ and $\sigma_2(X)$ in line 4 of Alg.~\ref{alg:plns}.

\begin{algorithm}[!htb]
\caption{repairSol}
\label{alg:repairSol}

\SetAlgoLined
\DontPrintSemicolon

\SetKwData{fP}{$\sigma_1$}
\SetKwData{fQ}{$\sigma_2$}

\KwIn{$X$, $B$, \fP, \fQ}

\BlankLine

$B' \leftarrow B$ \;
%\ShowLn
\lIf{$\fP < \fQ$}{ $i \leftarrow 1$ } \lElse{$i \leftarrow 2$} \label{algl:repairSol-choice}

\While{$X$ infeasible}{
$z^* \leftarrow \infty $, \quad $b^* \leftarrow \emptyset $ \; %\tcp*[r]{most greedy bid}

\ForEach{$b \in B' \setminus X$}{
	\uIf{$g_i(b,X) < z^*$}{
		$z^* \leftarrow g_i(b,X)$ \quad $b^* \leftarrow b$ \;
	}
  \ElseIf{$g_i(b,X) = \infty$}{$B' \leftarrow B' \setminus \{b\}$}
}

$X \leftarrow X \cup \{b^*\}$

}

\Return $X$ \;
\end{algorithm}

The heuristic decides in a self-adaptive manner which greedy function to use. The greedy function is chosen which produced least failures in generating new non-dominated solutions (cf. Alg.~\ref{alg:repairSol}, line~\ref{algl:repairSol-choice} and the calculation of $\sigma_1$ and $\sigma_2$ in Alg.~\ref{alg:plns}). Thereby, the procedure temporarily prefers one of the objective functions and tries to find a new solution which improves existing solutions with respect to the temporarily preferred objective function.
The remaining parts of ATlg.~\ref{alg:repairSol} form a straight forward single-criterion greedy procedure. Note, although the repair procedure implicitly uses an order of precedence for the objective functions, new solutions are solely accepted if they are non-dominated with respect to both objective function values (cf. Alg.~\ref{alg:construct}, line 13).

\subsection{Note on a matheuristic extension} \label{sec:alg_note}
\citet{Buer_2010a} introduced an exact branch-and-bound method based on the epsilon constraint approach for the 2WDP-SC. This approach, denoted as $\epsilon LBB$, was successfully used in \citet{Buer_2010b} to hybridize the path-relinking phase of a GRASP method for the 2WDP-SC. Obviously, we therefore also tried to further improve the solution quality of PNS by integrating $\epsilon LBB$ in three ways: 1) hybridizing $\epsilon$LBB and DRC, 2) hybridizing $\epsilon$LBB and PLNS, and 3) using $\epsilon$LBB within both, DRC and PLNS. All in all, given the same computing time, the three hybridized approaches led to inferior results compared to PNS \citep[p. 175f]{Buer_2012b}. It was difficult to define a subproblem of an appropriate size, either the subproblem was too small and its solution became trivial or the problem was too large to be solved by $\epsilon$LBB fast enough. Therefore, we did not pursue this research direction further.

\section{Design of computational study}
\label{sec:study}
The performance of the proposed heuristic is measured by means of a computational study. This section gives remarks on the test procedure, presents the benchmark instances used, and introduces measures for the quality of an approximation set from the literature.

\subsection{General remarks and test procedure} \label{sec:test}
The computational evaluation is done by means of artificial benchmark instances. All algorithms were implemented in Java (JDK 6, Update 23).
All tests were executed on the same type of personal computer (CPU Intel core i5-750, four cores, each 2.66 GHz). This also includes those heuristics that were published previously (cf. Sect. \ref{sec:test_comparison}), that is, previous computational experiments were repeated if necessary.
The tested algorithms are implemented sequentially and do not exploit features of multi-core processors.
However, in order to reduce the expenditure of time we needed for the evaluation of our algorithms, we perform four test runs at the same time on our  computer which has four cpu cores. This means, the test runs are not completely independent of each other, however, a significant negative interaction with respect to performance could not be observed.

We first evaluate some main design choices of the method PNS. At the same time, we work out reasonable values for the three external parameters of the heuristic PNS. Finally, the new method PNS is compared to three other heuristics from the literature.

\subsection{Benchmark instances} \label{sec:instances}
The 37 benchmark instances for the 2WDP-SC introduced in \cite{Buer_2010a} are used. They are open accessible via the electronic appendix of \cite{Buer_2010b}. These instances take into account some specific features of the transportation scenario at hand. In particular, the instance generation procedure creates bundle bids that satisfy the free disposal assumption. This is important, as this assumption was required to model the 2WDP-SC with set covering constraints (instead of set partitioning constraints).

For seven instances, all Pareto optimal solutions are known. These instances are denoted as \emph{small instances} (instance group S). The small instances feature up to 80 bundle bids. For the remaining thirty instances, the set of Pareto optimal solutions is unknown. These instances are denoted as  \emph{large instances}. These instances are divided into classes by means of different groups classifying instances according to their number of bundles or their number of contracts. There are three groups A, B, and C which contain instances with 500, 1000, and 2000 bundle bids, respectively. The groups a, b, c denote instances with 125, 250, and 500 transport contracts, respectively. Consequently, the class Cb for example contains those instances with 2000 bundle bids and 250 transport contracts. This notation is used in Tab. \ref{tab:results_S}, Tab. \ref{tab:results_ABC}, and Tab. \ref{tab:runtime}.

\subsection{Quality indicators for approximation sets} \label{sec:indicators}
The assessment of the quality of an approximation set is a nontrivial task. An intensive examination of different approaches is given by \citet{Zitzler_2003}.
The evaluation of algorithms in view of the obtained solution quality is usually more complex in the multiple objective case than in the single objective case.
In the single objective case, performance statements are naturally made by comparing the objective function values of solutions generated by different algorithms. However, in the multi objective case, approximation sets have to be compared whose fronts cross each other. Given two approximation sets $A$ and $B$ with solutions in $A$ that dominate solutions in $B$ and the other way round ($a \prec b$ and $b' \prec a'$ for $a, a' \in A$ and $b, b' \in B$) makes performance comparisons difficult.

One way to measure approximation set quality is the usage of quality indicators which should narrow down the comparison of two approximation sets to the comparison of two real-valued numbers. Roughly speaking, a quality indicator is a function that assigns to one or more approximation sets a scalar value. This always goes along with a loss of information intrinsic to the approximation sets. Likewise, if an approximation set $A$ outperforms an approximation set $B$ according to one indicator, one cannot conclude that each solution in $B$ is dominated by at least one solution in $A$.
Hence, it is advisable to use more than one quality indicator to balance the individual strengths and weaknesses of indicators (which are discussed e.g. by \citet{Zitzler_2003}). Therefore, we use three quality indicators for the computational study, the hypervolume indicator, the multiplicative epsilon indicator, and the coverage indicator. Those quality indicators seem to be among the most accepted and widely used in the literature.

\subsubsection{Hypervolume indicator $I_{HV}$}
\begin{figure}[!htbp]
	\centering
	\includegraphics[scale=0.6]{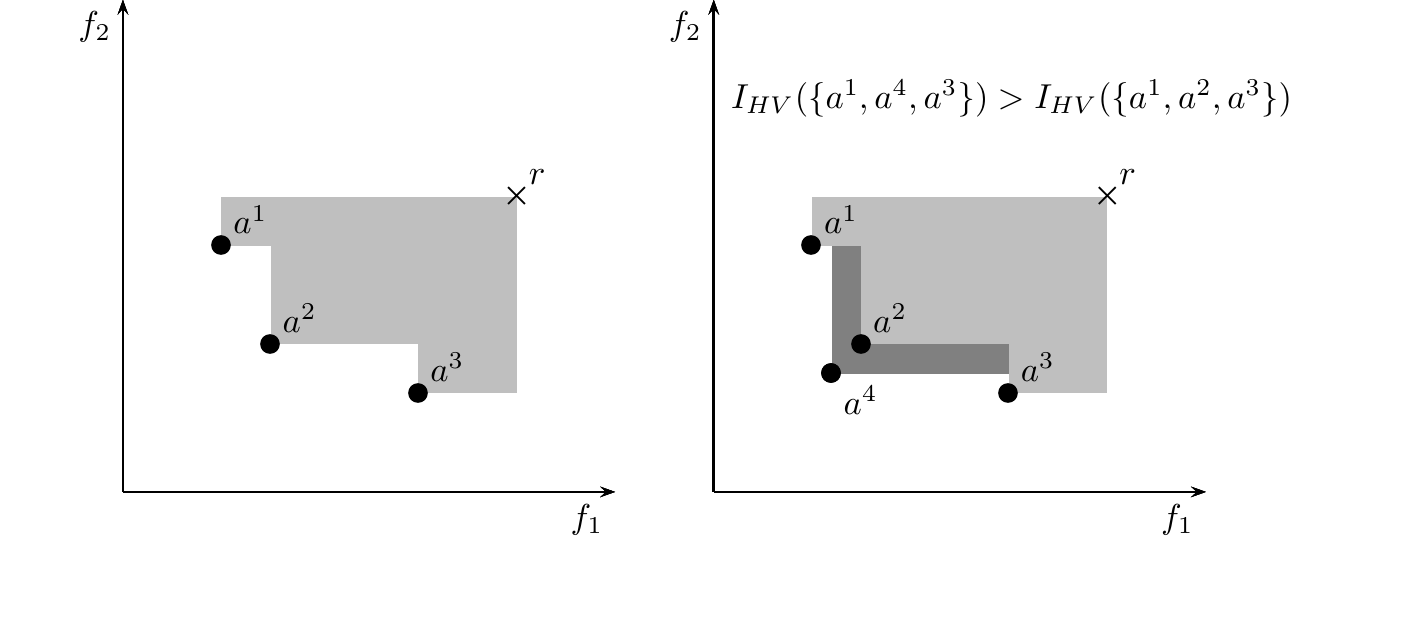}
	\caption{Principle of hypervolume indicator $I_{HV}$.}
	\label{fig:hypervolume}
\end{figure}

The hypervolume indicator $I_{HV}(A)$ measures the volume of the objective subspace that is weakly dominated by the solutions of a given approximation set $A$ and bounded by a reference point $r$ \citep{Zitzler_1998, Zitzler_1999}. The reference point $r$ has to be weakly dominated by each solution. Higher indicator values imply a better approximation set.
Fig. \ref{fig:hypervolume} (left) shows three non-dominated solutions $a^1, a^2, a^3$. The part of the objective space that is dominated by these solutions and bounded by the reference point $r$ is shaded in gray. The volume of the gray area is the value of $I_{HV}(\{a^1, a^2, a^3\})$. In Fig. \ref{fig:hypervolume} (right) the new non-dominated solution $a^4$ is added and the hypervolume is increased by the volume of the area shaded in dark gray. Apparently, every new non-dominated solution increases the value of $I_{HV}$.

In line with earlier studies on the 2WDP-SC \citep{Buer_2010a, Buer_2010b}, the reference point $r$ is defined as $r_1 = f_1(B)$ and $r_2 = 0$. The values of the objective functions $f_1$ and $f_2$ differ in several orders of magnitude (using the benchmark instances of Section \ref{sec:instances}). Therefore they are normalized prior to calculating $I_{HV}$ according to equation (\ref{eq:normalizing}).

\begin{align}
\label{eq:normalizing}
f'_i(X) = \frac{f_i(X) - f^{min}_{i}}{r_i - f^{min}_i} \quad \text{with} \;\; i \in \{1,2\}
\end{align}
and $f^{min}_1 := 0, \; f^{min}_2 := f_2(B)-1$.

\subsubsection{Epsilon indicator $I_{\epsilon}$}
The multiplicative epsilon indicator $I_{\epsilon}(A,B)$ introduced by \citet[p.~122]{Zitzler_2003} compares two approximation sets $A$ and $B$ and is based on the epsilon dominance relation $\preceq_{\epsilon}$. It is defined as follows:

\begin{align}
	f(a) \preceq_{\epsilon} f(b) \Leftrightarrow
	\forall i \in \{1, \ldots m \}: f_i(a) \leq \epsilon \cdot f_i(b).
\end{align}

$I_{\epsilon}(A,B)$ is the minimum factor, by which the value of the objective function of each solution in $B$ has to be multiplied, such that each solution in $B$ is epsilon dominated by at least one solution in $A$.

\begin{align}
	I_{\epsilon}(A,B) = \inf_{\epsilon \in \mathbb{R}}
	\{ \forall b \in B,\; \exists a \in A : f(a) \preceq_{\epsilon} f(b) \}.
\end{align}

Lower values of $I_{\epsilon}(A,B)$ imply a higher quality of $A$. By definition, it holds that $I_{\epsilon}(A,B) \geq 1$. For $I_{\epsilon}(A,B) = 1$, each solution in $B$ is weakly dominated by a solution in $A$. In general, $I_{\epsilon}(A,B) \neq I_{\epsilon}(B,A)$ holds.

$I_{\epsilon}(A,B)$ is a binary indicator. In case that more than two approximation sets should be compared, a pairwise comparison of the involved approximation sets is required. To simplify the comparison, in this study, we use the unary epsilon indicator \citep[S.~12]{Knowles_2006}:

\begin{align}
I_{\epsilon}(A) := I_{\epsilon}(A,A^R).
\end{align}

$A^R$ is denoted as reference approximation set. $A^R$ is the set union of the approximation sets $A'$ to be compared without any dominated solutions.

\subsubsection{Coverage indicator $I_C$}
\citet[S.~297]{Zitzler_1998} introduced the binary coverage indicator.
The coverage indicator $I_C(A,B)$ indicates the fraction of solutions in the approximation set $B$, that are dominated by at least one solution in the approximation set $A$.

\begin{align}
	I_C(A,B) = \frac{|\{b | \exists a \in A : f(a) \preceq f(b) \}|}{|B|}.
\end{align}

In general, $I_{\epsilon}(A,B) \neq I_{\epsilon}(B,A)$ holds. Higher values of $I_C(A,B)$ imply a higher quality of $A$. The range of values is $0 \leq I_C(A,B) \leq 1$, where $I_C(A,B)=1$ indicates that each solution in $B$ is dominated by at least one solution in $A$. Like $I_{\epsilon}(A, B)$, $I_C(A,B)$ is again a binary indicator and we only use the unary variant by means of a reference approximation set: $I_{C}(A) := I_{C}(A,A^R)$.

\section{Results and discussion} \label{sec:results}
\subsection{Contribution of two-stage candidate bid selection} \label{sec:res_construction}
We evaluate whether the quality of the approximation sets generated by the two-stage candidate bid selection procedure is improved in comparison to a traditional single-stage bid selection procedure. For this, only the construction procedure DRC (cf. Alg.~\ref{alg:construct}) is studied.
The single-stage selection procedure is realized by replacing the lines 1--11 of Alg.~\ref{alg:construct} with the single statement $C_i \leftarrow C$. The two-stage procedure is realized by DRC using multiple sectors ($s \geq 1$).

We first try to gain an insight into the actual size of the candidate list to identify a reasonable number of sectors $s$. The heuristic DRC (cf. Alg.~\ref{alg:construct}) computed 500 solutions ($l^{max} = \infty$) for each of the 37 instances.
Immediately prior to each call of the method \emph{selCandSector} in DRC, the size of the candidate list $C$ was logged. Tab.~\ref{tab:candidatelist} shows the results.
The average size of the candidate list $C$ grows slightly with increasing numbers of bundle bids per instance.
Nevertheless, even for the largest instances with 2000 bids, the median of $|C|$ is only 4 and the maximum size is 21.
To avoid an insufficient small number of bids per sector, we use three sectors ($s=3$) in the two-stage bid selection approach.

\begin{table}[!htbp]
\caption{Size of the candidate list $C$ during DRC.}
\label{tab:candidatelist}
\centering
\small
		\begin{tabular}{lcccc} \toprule
		Instance group & Mean & Stand. dev. & Median & Max. \\ \midrule
		S ($< 100$ bids)& 3.00 &  1.47 & 3 & 7 \\
		A (500 bids)& 4.19 & 2.22 & 4  & 16 \\
		B (1000 bids) & 4.29 & 2.51 & 4 & 16 \\
		C (2000 bids) & 4.58 & 2.91 & 4 & 21 \\ 		\bottomrule
			
		\end{tabular}
\\[0.5em]

\end{table}

Now we compare the single-stage and the two-stage approach.
The construction heuristic with a single-stage bid selection is denoted as DRC$_{s=1}$ and the two-stage bid selection heuristic is denoted as DRC$_{s=3}$.
In contrast to DRC (Alg. \ref{alg:construct}), both heuristics terminate after 1000 constructed solutions (and $l^{max} = \infty$).
For each of the thirty large instances five test runs with different random seeds were performed. A test run is the one-time computation of an instance with both heuristics DRC$_{s=1}$ and DRC$_{s=3}$.
The results for the quality indicators $I_{HV}$, $I_{\epsilon}$, and $I_{C}$ are shown in Tab.~\ref{tab:res_CL}. The rows $Q_{25}$, $Q_{50}$, and $Q_{75}$ show the lower quartile, the median, and the upper quartile, respectively, of the 150 resulting indicator values.
Note, each test run leads to its own approximation set $A^R$. Hence, the solutions generated during multiple test runs on the same instance are not mixed up to generate the approximation set.

\begin{table}[!htbp]
\caption{One-stage (DRC$_{s=1}$) versus two-stage (DRC$_{s=3}$) selection of bids from the candidate list.}
\label{tab:res_CL}
\centering
\small{

\begin{tabular}{C{0.45cm}llllll} \toprule

& \multicolumn{2}{c}{$I_{HV}$} & \multicolumn{2}{c}{$I_{\epsilon}$} & \multicolumn{2}{c}{$I_{C}$} \\

\cmidrule(lr){2-3} \cmidrule(lr){4-5} \cmidrule(lr){6-7}

& \multicolumn{1}{R{0.6cm}}{\tiny{DRC$_{s=1}$}} & \multicolumn{1}{R{0.6cm}}{\tiny{DRC$_{s=3}$}} & 	\multicolumn{1}{R{0.6cm}}{\tiny{DRC$_{s=1}$}} & \multicolumn{1}{R{0.6cm}}{\tiny{DRC$_{s=3}$}} & \multicolumn{1}{R{0.6cm}}{\tiny{DRC$_{s=1}$}} & \multicolumn{1}{R{0.6cm}}{\tiny{DRC$_{s=3}$}} \\ \midrule

$Q_{25}$ & \multicolumn{1}{l}{0.8815} & \multicolumn{1}{l}{0.8924} & \multicolumn{1}{l}{1.09} 	& \multicolumn{1}{l}{1.13} 	& \multicolumn{1}{l}{0.07} 	& \multicolumn{1}{l}{0.05} 	\\

$Q_{50}$ & \multicolumn{1}{l}{0.8917} & \multicolumn{1}{l}{0.8992} & \multicolumn{1}{l}{3} 	& \multicolumn{1}{l}{1.2} 	&  \multicolumn{1}{l}{0.11}	&  \multicolumn{1}{l}{0.09} 	\\

$Q_{75}$ & \multicolumn{1}{l}{0.9352} & \multicolumn{1}{l}{0.9526} & \multicolumn{1}{l}{13} 	& \multicolumn{1}{l}{1.29} 	&  \multicolumn{1}{l}{0.15} 	&  \multicolumn{1}{l}{0.13} 	\\ \bottomrule

\end{tabular}}
\end{table}

Applying the Wilcoxon signed rank test to the results, the null hypothesis ('the quality indicator median values of the different tested algorithms possess the same probability distribution') can be rejected for each of the three quality indicators. The p-values for $I_{HV}$, $I_{\epsilon}$, and $I_C$ are  $\leq$0.0001, $\leq$0.0001, and $0.0028$, respectively. The results are statistically significant even for very tight levels of significance of one percent or lower.
Therefore, it is highly likely, that the observed quality differences of the obtained approximation sets can be attributed to the usage of the two-stage bid selection procedure in the construction phase.

Regarding the median values of the quality indicators $I_{HV}$ and $I_{\epsilon}$, the two-stage approach DRC$_{s=3}$ clearly outperforms the one-stage heuristic DRC$_{s=1}$. In contrast, the median values of $I_C$ suggest an opposite interpretation. Looking at the generated approximation sets, the two-stage heuristic seems to discover better extreme solutions than $DRC_{s=1}$, especially with respect to $f_1$. On the other hand, DRC$_{s=1}$ seems to generate more and better compromise solutions with balanced $f_1$ and $f_2$ values. This might explain the slightly better values of $I_C$ in Tab. \ref{tab:res_CL}. All in all, the two-stage candidate bid selection approach generally increases the quality of the calculated approximation sets.

\subsection{Contribution of dynamic destroy rates} \label{sec:res_improvement}

The goal of the next experiment is twofold. On the one hand, we want to check if the proposed dynamic destroy rates in the improvement phase contribute to achieve higher approximation set quality. On the other hand, a proper destroy strategy is searched for. Recall, a destroy strategy $\bm d$ is a sequence $(d_1, \ldots, d_n)$ of destroy rates. We compare 17 destroy strategies which are indicated in the first column of Tab.~\ref{tab:res_destroystrategies}.
In configuring the destroy strategies, we tried to reflect several patterns: increasing destroy rates, decreasing destroy rates, and mixed variants. A higher destroy rate leads to a larger neighborhood which might enable the search to overcome local optima. However, the reconstruction of a solution is greedy which should favor smaller neighborhoods. Because of these opposed effects, different strategies were tested.
Furthermore, there are five static strategies $(3)$, $(6)$, $(9)$, $(12)$, $(15)$ whose destroy rate is constant and a self-adaptive choice between several destroy rates is impossible. We also experimented with larger destroy rates between 20 and 40 percent, however, these seem clearly inferior to strategies with smaller destroy rates reported in Tab. \ref{tab:res_destroystrategies}.
Each of  the 17 destroy strategies is used to compute the thirty large instances twice.
Column two to four of Tab.~\ref{tab:res_destroystrategies} show the median of the appropriate quality indicator over 60 runs of the destroy strategy. The runtime for each run was fixed to five minutes. The best median values are bold.

\begin{table}[!htbp]
\caption{Results for 17 destroy strategies.}
\centering
\label{tab:res_destroystrategies}
\small

\begin{tabular}{lllll} \toprule
$\bm d = (d_1, \ldots, d_n)$ & \multicolumn{1}{c}{$I_{HV}$} & \multicolumn{1}{c}{$I_{\epsilon}$} & \multicolumn{1}{c}{$I_{C}$} \\ \midrule
$(3, 6, 9)$ & 0.9095 & $\textbf{1.01}$ & 0.01 \\
$(6, 12, 18)$ & 0.9093 & 1.02 & 0.00 \\
$(9, 18, 27)$ & 0.9089 & 1.03 & 0.00 \\[0.3em]
$(9, 6, 3)$ & \textbf{0.9097} & 1.015 & 0.01 \\
$(18, 12, 6)$ & 0.9096 & 1.02 & 0.00 \\
$(27, 18, 9)$ & 0.9093 & 1.03 & 0.00 \\[0.3em]
$(3)$ & 0.9096 & \textbf{1.01} & \textbf{0.07} \\
$(6)$ & 0.9096 & 1.02 & 0.02 \\
$(9)$ & 0.9096 & 1.02 & 0.00 \\
$(12)$ & 0.9092 & 1.025 & 0.00 \\
$(15)$ & 0.9092 & 1.03 & 0.00 \\[0.3em]
$(5, 15, 7)$ & 0.9096 & 1.02 & 0.00 \\
$(7, 19, 9)$ & 0.9093 & 1.02 & 0.00 \\[0.3em]
$(15, 5, 10)$ & 0.9094 & 1.02 & 0.00 \\
$(19, 7, 14)$ & 0.9095 & 1.02 & 0.00 \\[0.3em]
$(3, 6, 9, 2, 4)$ & \textbf{0.9097} & \textbf{1.01} & 0.05 \\
$(6, 12, 18, 5, 10)$ & 0.9096 & 1.02 & 0.00 \\
\bottomrule
\end{tabular}
\\[0.5em]
Median values of quality indicators over two runs for each large instance.
\end{table}

The strategies $(3)$ and $(3, 6, 9, 2, 4)$ achieve the best median values for two quality indicators, respectively. To decide which is superior, we use both strategies to compute each of the large instances five times. Applying the Wilcoxon signed rank test to the results, the null hypothesis ('the quality indicator median values of the tested algorithms possess the same probability distribution') can be rejected for two of the three quality indicators on a level of significance of less than three percent. The p-values for $I_{HV}$, $I_{\epsilon}$, and $I_C$ are  $\leq$0.0001, 0.0216, and $0.4231$, respectively. The dynamic strategy $(3, 6, 9, 2, 4)$ clearly outperforms the static strategy $(3)$ by means of the hypervolume indicator and the epsilon indicator while the observed difference by means of the coverage indicator is not significant. We conclude, that the dynamic strategy $(3, 6, 9, 2, 4)$ works best. This conclusion is also supported by empirical runtime distributions, which are discussed in Fig. \ref{fig:nbs-runtime} in the appendix.

\subsection{Comparison with other heuristics}
\label{sec:test_comparison}
To benchmark the new method PNS by means of approximation set quality the three heuristics SPEA2A \cite{Buer_2010a}, PGRASP$_P$+HPR \cite{Buer_2010b}, and PGRASP$_Q$+HPR \cite{Buer_2010b} are used.
All three are problem-specific heuristics for the 2WDP-SC.
The method SPEA2A, however, is based on the \emph{Strength Pareto Evolutionary Algorithm 2} (SPEA2), an originally problem-independent heuristic introduced by \citet{Zitzler_2001}.
In \cite{Buer_2010a}, the 2WDP-SC was solved by eight variants of SPEA2 with different problem-specific construction, mutation, and repair operators.
The variant which performed best was denoted as $A_8$ in \cite{Buer_2010a}; in the present paper we denote this variant as SPEA2A.
As expected, variants which used more problem knowledge by means of their operators clearly outperformed those variants that used less problem knowledge.
Due to these results, we suppose the enhanced techniques to cope with multiple objectives used in recent problem-independent approaches like EMOSA or MOEA/D still cannot overcompensate the absence of problem knowledge during search.
Therefore, we avoid to include more recent but problem-independent approaches like EMOSA or MOEA/D and we compare PNS only to those heuristics from the literature that incorporate problem-specific knowledge during search.

PGRASP$_P$+HPR, and PGRASP$_Q$+HPR were proposed in \cite{Buer_2010b}.
Both methods are multi objective GRASP whose path-relinking phase was hybridized with the exact branch-and-bound method $\epsilon LBB$ of \cite{Buer_2010a}. Another hybridized heuristic for the 2WDP-SC was discussed in \cite{Buer_2011} (see also \ref{sec:alg_note}) which is, however, not included in our comparison, as it does not clearly outperform the mentioned heuristics on the majority of instances.

For the benchmark, the parameters of PNS are set as follows.
The number of sections $s$ are set to $3$, the vector destroy probabilities is set to (3,6,9,2,4), and the termination criterion of the construction phase is set to $l^{max}=92$.
While the configuration of the first two values were justified in Sections~\ref{sec:res_construction} and~\ref{sec:res_improvement}, the value of the termination criterion $l^{max}$ was determined as follows: for each large instance, 1000 solutions were generated with DRC. The experimental distribution of the number of unsuccessful improvement tries was recorded (median 6, mean 20, standard deviation 42) and $l^{max}$ was set to the value of the ninety-five percent quantile, which is 92.

The runtime of each heuristic was five minutes (300s). All heuristics computed all instances on the same type of computer.
Please note, we \emph{do not cite} the computational results of the experiments in \cite{Buer_2010a, Buer_2010b} but compute all instances again on a faster computer.

The results for the \emph{small instances} (group S) with known Pareto optimal solution sets are shown in Tab.~\ref{tab:results_S}.
The two rightmost columns show the Pareto optimal hypervolume values and the cardinality of the reference approximation set $A^R$ (here, it is identical to the Pareto optimum solution set). These optimal results haven been obtained by the bicriteria branch-and-bound method $\epsilon LBB$ introduced in \cite{Buer_2010a}.

Algorithm PNS is able to solve all seven small instances to Pareto optimality, that is the whole Pareto optimal solution set is found. In contrast, the procedures PGRASP$_P$+HPR and PGRASP$_Q$+HPR are able to solve six out of seven instances to Pareto optimality.
In \citep{Buer_2010b}, only four instances could be solved to Pareto optimality.
The method SPEA2A is able to find some Pareto optimal solutions for six instances (S1 -- S5, S7), but never the complete set.

\begin{table*}[!htbp]
	\caption{Comparison of solution approaches by means of small instances (instance group S).}
	\label{tab:results_S}
	\centering
	\scriptsize

\begin{tabular}{lllllllllllllll} \toprule
Instance & \multicolumn{3}{c}{PNS} & \multicolumn{3}{c}{PGRASP$_P$+HPR} & \multicolumn{3}{c}{PGRASP$_Q$+HPR} & \multicolumn{3}{c}{SPEA2A} & \multicolumn{2}{c}{$\epsilon$LBB*} \\
\cmidrule(lr){2-4} \cmidrule(lr){5-7} \cmidrule(lr){8-10} \cmidrule(lr){11-13} \cmidrule(lr){14-15}
 & \multicolumn{1}{c}{$I_{HV}$} & \multicolumn{1}{c}{$I_{\epsilon}$} & \multicolumn{1}{c}{$I_{C}$} & \multicolumn{1}{c}{$I_{HV}$} & \multicolumn{1}{c}{$I_{\epsilon}$} & \multicolumn{1}{c}{$I_{C}$} & \multicolumn{1}{c}{$I_{HV}$} & \multicolumn{1}{c}{$I_{\epsilon}$} & \multicolumn{1}{c}{$I_{C}$} & \multicolumn{1}{c}{$I_{HV}$} & \multicolumn{1}{c}{$I_{\epsilon}$} & \multicolumn{1}{c}{$I_{C}$} & \multicolumn{1}{c}{$I_{HV}$} & \multicolumn{1}{c}{$|A^R|$} \\ \midrule
S1 & \multicolumn{1}{r}{0.8576} & \multicolumn{1}{r}{1.00} & \multicolumn{1}{r}{1.00} & \multicolumn{1}{r}{0.8576} & \multicolumn{1}{r}{1.00} & \multicolumn{1}{r}{1.00} & \multicolumn{1}{r}{0.8576} & \multicolumn{1}{r}{1.00} & \multicolumn{1}{r}{1.00} & \multicolumn{1}{r}{0.8573} & \multicolumn{1}{r}{1.03} & \multicolumn{1}{r}{0.71} & \multicolumn{1}{r}{0.8576} & \multicolumn{1}{r}{7} \\
S2 & \multicolumn{1}{r}{0.6095} & \multicolumn{1}{r}{1.00} & \multicolumn{1}{r}{1.00} & \multicolumn{1}{r}{0.6095} & \multicolumn{1}{r}{1.00} & \multicolumn{1}{r}{1.00} & \multicolumn{1}{r}{0.6095} & \multicolumn{1}{r}{1.00} & \multicolumn{1}{r}{1.00} & \multicolumn{1}{r}{0.6022} & \multicolumn{1}{r}{1.08} & \multicolumn{1}{r}{0.45} & \multicolumn{1}{r}{0.6095} & \multicolumn{1}{r}{11} \\
S3 & \multicolumn{1}{r}{0.8169} & \multicolumn{1}{r}{1.00} & \multicolumn{1}{r}{1.00} & \multicolumn{1}{r}{0.8169} & \multicolumn{1}{r}{1.00} & \multicolumn{1}{r}{1.00} & \multicolumn{1}{r}{0.8169} & \multicolumn{1}{r}{1.00} & \multicolumn{1}{r}{1.00} & \multicolumn{1}{r}{0.8125} & \multicolumn{1}{r}{1.47} & \multicolumn{1}{r}{0.38} & \multicolumn{1}{r}{0.8169} & \multicolumn{1}{r}{13} \\
S4 & \multicolumn{1}{r}{0.5677} & \multicolumn{1}{r}{1.00} & \multicolumn{1}{r}{1.00} & \multicolumn{1}{r}{0.5677} & \multicolumn{1}{r}{1.00} & \multicolumn{1}{r}{1.00} & \multicolumn{1}{r}{0.5677} & \multicolumn{1}{r}{1.00} & \multicolumn{1}{r}{1.00} & \multicolumn{1}{r}{0.5636} & \multicolumn{1}{r}{1.41} & \multicolumn{1}{r}{0.25} & \multicolumn{1}{r}{0.5677} & \multicolumn{1}{r}{12} \\
S5 & \multicolumn{1}{r}{0.8652} & \multicolumn{1}{r}{1.00} & \multicolumn{1}{r}{1.00} & \multicolumn{1}{r}{0.8644} & \multicolumn{1}{r}{1.01} & \multicolumn{1}{r}{0.88} & \multicolumn{1}{r}{0.8652} & \multicolumn{1}{r}{1.02} & \multicolumn{1}{r}{0.94} & \multicolumn{1}{r}{0.8535} & \multicolumn{1}{r}{2.00} & \multicolumn{1}{r}{0.29} & \multicolumn{1}{r}{0.8652} & \multicolumn{1}{r}{17} \\
S6 & \multicolumn{1}{r}{0.6988} & \multicolumn{1}{r}{1.00} & \multicolumn{1}{r}{1.00} & \multicolumn{1}{r}{0.6988} & \multicolumn{1}{r}{1.00} & \multicolumn{1}{r}{1.00} & \multicolumn{1}{r}{0.6988} & \multicolumn{1}{r}{1.00} & \multicolumn{1}{r}{1.00} & \multicolumn{1}{r}{0.6879} & \multicolumn{1}{r}{1.27} & \multicolumn{1}{r}{0.10} & \multicolumn{1}{r}{0.6988} & \multicolumn{1}{r}{10} \\
S7 & \multicolumn{1}{r}{0.8915} & \multicolumn{1}{r}{1.00} & \multicolumn{1}{r}{1.00} & \multicolumn{1}{r}{0.8915} & \multicolumn{1}{r}{1.00} & \multicolumn{1}{r}{1.00} & \multicolumn{1}{r}{0.8915} & \multicolumn{1}{r}{1.00} & \multicolumn{1}{r}{1.00} & \multicolumn{1}{r}{0.8866} & \multicolumn{1}{r}{1.66} & \multicolumn{1}{r}{0.12} & \multicolumn{1}{r}{0.8915} & \multicolumn{1}{r}{17} \\ \bottomrule
\end{tabular}
\\[0.5em]
*The method $\epsilon$LBB calculates always Pareto-optimal solutions.
\end{table*}

The results for the \emph{large instances} (groups A, B, and C) without known Pareto optimum solutions are shown in Tab. \ref{tab:results_ABC}. This time, the reference approximation set $A^R$ (cf. two rightmost columns) is generated by merging the approximation sets of PNS, PGRASP$_P$+HPR, PGRASP$_Q$+HPR, and SPEA2A and removing the dominated solutions. The last five rows of Tab. \ref{tab:results_ABC} show the 25 percent quantile, the median, the 75 percent quantile, the mean, and the standard deviation for each heuristic and each quality indicator.

In terms of approximation set quality, all existing approaches  are clearly outperformed by the heuristic PNS.
The heuristic PNS finds new best approximation sets in terms of $I_{HV}$ and $I_{\epsilon}$ for all thirty instances.
Furthermore, from the values of $I_C$ it follows that the approximation sets computed by PNS are equal to the reference approximation set $A^R$ in 28 out of 30 instances.
That is, PNS dominates \emph{each} solution found by one of the benchmark heuristics.
Only for the instances Ba3 and Cc7, the reference approximation set is not solely generated by PNS.
Consequently, the solution approach PNS obtains for all three quality indicators the best median indicator values at the same time.

\begin{table*}[!htbp]
\caption{Comparison of solution approaches by means of large instances (instance groups A, B, and C).}
\label{tab:results_ABC}
\centering
\scriptsize

\begin{tabular}{lllllllllllllll} \toprule
Instance & \multicolumn{3}{c}{PNS} & \multicolumn{3}{c}{PGRASP$_P$+HPR} & \multicolumn{3}{c}{PGRASP$_Q$+HPR} & \multicolumn{3}{c}{SPEA2A} & \multicolumn{2}{c}{reference} \\ \cmidrule(lr){2-4}\cmidrule(lr){5-7}\cmidrule(lr){8-10}\cmidrule(lr){11-13}\cmidrule(lr){14-15}

 & \multicolumn{1}{c}{$I_{HV}$} & \multicolumn{1}{c}{$I_{\epsilon}$} & \multicolumn{1}{c}{$I_{C}$} & \multicolumn{1}{c}{$I_{HV}$} & \multicolumn{1}{c}{$I_{\epsilon}$} & \multicolumn{1}{c}{$I_{C}$} & \multicolumn{1}{c}{$I_{HV}$} & \multicolumn{1}{c}{$I_{\epsilon}$} & \multicolumn{1}{c}{$I_{C}$} & \multicolumn{1}{c}{$I_{HV}$} & \multicolumn{1}{c}{$I_{\epsilon}$} & \multicolumn{1}{c}{$I_{C}$} & \multicolumn{1}{c}{$I_{HV}$} & \multicolumn{1}{c}{$|A^R|$} \\
 \midrule

Aa1 & \multicolumn{1}{r}{0.9027} & \multicolumn{1}{r}{1.00} & \multicolumn{1}{r}{1.00} & \multicolumn{1}{r}{0.8996} & \multicolumn{1}{r}{1.14} & \multicolumn{1}{r}{0.00} & \multicolumn{1}{r}{0.8929} & \multicolumn{1}{r}{1.11} & \multicolumn{1}{r}{0.00} & \multicolumn{1}{r}{0.8895} & \multicolumn{1}{r}{1.33} & \multicolumn{1}{r}{0.00} & \multicolumn{1}{r}{0.9027} & \multicolumn{1}{r}{68} \\
Aa2 & \multicolumn{1}{r}{0.9132} & \multicolumn{1}{r}{1.00} & \multicolumn{1}{r}{1.00} & \multicolumn{1}{r}{0.9118} & \multicolumn{1}{r}{1.06} & \multicolumn{1}{r}{0.00} & \multicolumn{1}{r}{0.9056} & \multicolumn{1}{r}{1.09} & \multicolumn{1}{r}{0.00} & \multicolumn{1}{r}{0.9016} & \multicolumn{1}{r}{1.42} & \multicolumn{1}{r}{0.00} & \multicolumn{1}{r}{0.9132} & \multicolumn{1}{r}{43} \\
Aa3 & \multicolumn{1}{r}{0.9063} & \multicolumn{1}{r}{1.00} & \multicolumn{1}{r}{1.00} & \multicolumn{1}{r}{0.9026} & \multicolumn{1}{r}{1.04} & \multicolumn{1}{r}{0.00} & \multicolumn{1}{r}{0.8996} & \multicolumn{1}{r}{1.08} & \multicolumn{1}{r}{0.00} & \multicolumn{1}{r}{0.8979} & \multicolumn{1}{r}{1.29} & \multicolumn{1}{r}{0.00} & \multicolumn{1}{r}{0.9063} & \multicolumn{1}{r}{60} \\ [0.5em]
Ba1 & \multicolumn{1}{r}{0.9559} & \multicolumn{1}{r}{1.00} & \multicolumn{1}{r}{1.00} & \multicolumn{1}{r}{0.9521} & \multicolumn{1}{r}{1.21} & \multicolumn{1}{r}{0.00} & \multicolumn{1}{r}{0.9510} & \multicolumn{1}{r}{1.13} & \multicolumn{1}{r}{0.00} & \multicolumn{1}{r}{0.9475} & \multicolumn{1}{r}{1.43} & \multicolumn{1}{r}{0.00} & \multicolumn{1}{r}{0.9559} & \multicolumn{1}{r}{100} \\
Ba2 & \multicolumn{1}{r}{0.9596} & \multicolumn{1}{r}{1.00} & \multicolumn{1}{r}{1.00} & \multicolumn{1}{r}{0.9578} & \multicolumn{1}{r}{1.11} & \multicolumn{1}{r}{0.00} & \multicolumn{1}{r}{0.9545} & \multicolumn{1}{r}{1.14} & \multicolumn{1}{r}{0.00} & \multicolumn{1}{r}{0.9502} & \multicolumn{1}{r}{1.85} & \multicolumn{1}{r}{0.00} & \multicolumn{1}{r}{0.9596} & \multicolumn{1}{r}{80} \\
Ba3 & \multicolumn{1}{r}{0.9619} & \multicolumn{1}{r}{1.00} & \multicolumn{1}{r}{0.99} & \multicolumn{1}{r}{0.9595} & \multicolumn{1}{r}{1.22} & \multicolumn{1}{r}{0.00} & \multicolumn{1}{r}{0.9557} & \multicolumn{1}{r}{1.19} & \multicolumn{1}{r}{0.01} & \multicolumn{1}{r}{0.9521} & \multicolumn{1}{r}{1.36} & \multicolumn{1}{r}{0.00} & \multicolumn{1}{r}{0.9619} & \multicolumn{1}{r}{70} \\ [0.5em]
Bb1 & \multicolumn{1}{r}{0.9084} & \multicolumn{1}{r}{1.00} & \multicolumn{1}{r}{1.00} & \multicolumn{1}{r}{0.9050} & \multicolumn{1}{r}{1.17} & \multicolumn{1}{r}{0.00} & \multicolumn{1}{r}{0.9013} & \multicolumn{1}{r}{1.08} & \multicolumn{1}{r}{0.00} & \multicolumn{1}{r}{0.8971} & \multicolumn{1}{r}{1.34} & \multicolumn{1}{r}{0.00} & \multicolumn{1}{r}{0.9084} & \multicolumn{1}{r}{58} \\
Bb2 & \multicolumn{1}{r}{0.9070} & \multicolumn{1}{r}{1.00} & \multicolumn{1}{r}{1.00} & \multicolumn{1}{r}{0.9036} & \multicolumn{1}{r}{1.13} & \multicolumn{1}{r}{0.00} & \multicolumn{1}{r}{0.9009} & \multicolumn{1}{r}{1.07} & \multicolumn{1}{r}{0.00} & \multicolumn{1}{r}{0.8990} & \multicolumn{1}{r}{1.29} & \multicolumn{1}{r}{0.00} & \multicolumn{1}{r}{0.9070} & \multicolumn{1}{r}{65} \\
Bb3 & \multicolumn{1}{r}{0.9050} & \multicolumn{1}{r}{1.00} & \multicolumn{1}{r}{1.00} & \multicolumn{1}{r}{0.9033} & \multicolumn{1}{r}{1.14} & \multicolumn{1}{r}{0.00} & \multicolumn{1}{r}{0.8984} & \multicolumn{1}{r}{1.07} & \multicolumn{1}{r}{0.00} & \multicolumn{1}{r}{0.8960} & \multicolumn{1}{r}{1.33} & \multicolumn{1}{r}{0.00} & \multicolumn{1}{r}{0.9050} & \multicolumn{1}{r}{51} \\
Bb4 & \multicolumn{1}{r}{0.9143} & \multicolumn{1}{r}{1.00} & \multicolumn{1}{r}{1.00} & \multicolumn{1}{r}{0.9071} & \multicolumn{1}{r}{1.36} & \multicolumn{1}{r}{0.00} & \multicolumn{1}{r}{0.9024} & \multicolumn{1}{r}{2.00} & \multicolumn{1}{r}{0.00} & \multicolumn{1}{r}{0.8840} & \multicolumn{1}{r}{20.00} & \multicolumn{1}{r}{0.00} & \multicolumn{1}{r}{0.9143} & \multicolumn{1}{r}{149} \\
Bb5 & \multicolumn{1}{r}{0.9071} & \multicolumn{1}{r}{1.00} & \multicolumn{1}{r}{1.00} & \multicolumn{1}{r}{0.9006} & \multicolumn{1}{r}{1.12} & \multicolumn{1}{r}{0.00} & \multicolumn{1}{r}{0.8988} & \multicolumn{1}{r}{1.10} & \multicolumn{1}{r}{0.00} & \multicolumn{1}{r}{0.8913} & \multicolumn{1}{r}{2.00} & \multicolumn{1}{r}{0.00} & \multicolumn{1}{r}{0.9071} & \multicolumn{1}{r}{108} \\
Bb6 & \multicolumn{1}{r}{0.9102} & \multicolumn{1}{r}{1.00} & \multicolumn{1}{r}{1.00} & \multicolumn{1}{r}{0.9041} & \multicolumn{1}{r}{1.24} & \multicolumn{1}{r}{0.00} & \multicolumn{1}{r}{0.8993} & \multicolumn{1}{r}{1.13} & \multicolumn{1}{r}{0.00} & \multicolumn{1}{r}{0.8941} & \multicolumn{1}{r}{1.35} & \multicolumn{1}{r}{0.00} & \multicolumn{1}{r}{0.9102} & \multicolumn{1}{r}{114} \\ [0.5em]
Ca1 & \multicolumn{1}{r}{0.9809} & \multicolumn{1}{r}{1.00} & \multicolumn{1}{r}{1.00} & \multicolumn{1}{r}{0.9798} & \multicolumn{1}{r}{1.21} & \multicolumn{1}{r}{0.00} & \multicolumn{1}{r}{0.9783} & \multicolumn{1}{r}{1.18} & \multicolumn{1}{r}{0.00} & \multicolumn{1}{r}{0.9579} & \multicolumn{1}{r}{11.00} & \multicolumn{1}{r}{0.00} & \multicolumn{1}{r}{0.9809} & \multicolumn{1}{r}{100} \\
Ca2 & \multicolumn{1}{r}{0.9825} & \multicolumn{1}{r}{1.00} & \multicolumn{1}{r}{1.00} & \multicolumn{1}{r}{0.9809} & \multicolumn{1}{r}{1.35} & \multicolumn{1}{r}{0.00} & \multicolumn{1}{r}{0.9794} & \multicolumn{1}{r}{1.22} & \multicolumn{1}{r}{0.00} & \multicolumn{1}{r}{0.9793} & \multicolumn{1}{r}{1.53} & \multicolumn{1}{r}{0.00} & \multicolumn{1}{r}{0.9825} & \multicolumn{1}{r}{85} \\
Ca3 & \multicolumn{1}{r}{0.9812} & \multicolumn{1}{r}{1.00} & \multicolumn{1}{r}{1.00} & \multicolumn{1}{r}{0.9781} & \multicolumn{1}{r}{2.00} & \multicolumn{1}{r}{0.00} & \multicolumn{1}{r}{0.9786} & \multicolumn{1}{r}{1.17} & \multicolumn{1}{r}{0.00} & \multicolumn{1}{r}{0.9691} & \multicolumn{1}{r}{5.00} & \multicolumn{1}{r}{0.00} & \multicolumn{1}{r}{0.9812} & \multicolumn{1}{r}{73} \\ [0.5em]
Cb1 & \multicolumn{1}{r}{0.9585} & \multicolumn{1}{r}{1.00} & \multicolumn{1}{r}{1.00} & \multicolumn{1}{r}{0.9560} & \multicolumn{1}{r}{1.15} & \multicolumn{1}{r}{0.00} & \multicolumn{1}{r}{0.9529} & \multicolumn{1}{r}{1.15} & \multicolumn{1}{r}{0.00} & \multicolumn{1}{r}{0.9527} & \multicolumn{1}{r}{1.27} & \multicolumn{1}{r}{0.00} & \multicolumn{1}{r}{0.9585} & \multicolumn{1}{r}{78} \\
Cb2 & \multicolumn{1}{r}{0.9589} & \multicolumn{1}{r}{1.00} & \multicolumn{1}{r}{1.00} & \multicolumn{1}{r}{0.9567} & \multicolumn{1}{r}{1.21} & \multicolumn{1}{r}{0.00} & \multicolumn{1}{r}{0.9539} & \multicolumn{1}{r}{1.13} & \multicolumn{1}{r}{0.00} & \multicolumn{1}{r}{0.9527} & \multicolumn{1}{r}{1.35} & \multicolumn{1}{r}{0.00} & \multicolumn{1}{r}{0.9589} & \multicolumn{1}{r}{50} \\
Cb3 & \multicolumn{1}{r}{0.9569} & \multicolumn{1}{r}{1.00} & \multicolumn{1}{r}{1.00} & \multicolumn{1}{r}{0.9544} & \multicolumn{1}{r}{1.09} & \multicolumn{1}{r}{0.00} & \multicolumn{1}{r}{0.9530} & \multicolumn{1}{r}{1.09} & \multicolumn{1}{r}{0.00} & \multicolumn{1}{r}{0.9512} & \multicolumn{1}{r}{1.32} & \multicolumn{1}{r}{0.00} & \multicolumn{1}{r}{0.9569} & \multicolumn{1}{r}{30} \\
Cb4 & \multicolumn{1}{r}{0.9594} & \multicolumn{1}{r}{1.00} & \multicolumn{1}{r}{1.00} & \multicolumn{1}{r}{0.9546} & \multicolumn{1}{r}{2.00} & \multicolumn{1}{r}{0.00} & \multicolumn{1}{r}{0.9533} & \multicolumn{1}{r}{1.17} & \multicolumn{1}{r}{0.00} & \multicolumn{1}{r}{0.9410} & \multicolumn{1}{r}{13.00} & \multicolumn{1}{r}{0.00} & \multicolumn{1}{r}{0.9594} & \multicolumn{1}{r}{143} \\
Cb5 & \multicolumn{1}{r}{0.9621} & \multicolumn{1}{r}{1.00} & \multicolumn{1}{r}{1.00} & \multicolumn{1}{r}{0.9581} & \multicolumn{1}{r}{1.42} & \multicolumn{1}{r}{0.00} & \multicolumn{1}{r}{0.9556} & \multicolumn{1}{r}{1.19} & \multicolumn{1}{r}{0.00} & \multicolumn{1}{r}{0.9495} & \multicolumn{1}{r}{7.00} & \multicolumn{1}{r}{0.00} & \multicolumn{1}{r}{0.9621} & \multicolumn{1}{r}{119} \\
Cb6 & \multicolumn{1}{r}{0.9586} & \multicolumn{1}{r}{1.00} & \multicolumn{1}{r}{1.00} & \multicolumn{1}{r}{0.9537} & \multicolumn{1}{r}{1.33} & \multicolumn{1}{r}{0.00} & \multicolumn{1}{r}{0.9524} & \multicolumn{1}{r}{1.17} & \multicolumn{1}{r}{0.00} & \multicolumn{1}{r}{0.9507} & \multicolumn{1}{r}{1.54} & \multicolumn{1}{r}{0.00} & \multicolumn{1}{r}{0.9586} & \multicolumn{1}{r}{100} \\  [0.5em]
Cc1 & \multicolumn{1}{r}{0.8991} & \multicolumn{1}{r}{1.00} & \multicolumn{1}{r}{1.00} & \multicolumn{1}{r}{0.8914} & \multicolumn{1}{r}{2.00} & \multicolumn{1}{r}{0.00} & \multicolumn{1}{r}{0.8883} & \multicolumn{1}{r}{1.11} & \multicolumn{1}{r}{0.00} & \multicolumn{1}{r}{0.8773} & \multicolumn{1}{r}{27.00} & \multicolumn{1}{r}{0.00} & \multicolumn{1}{r}{0.8991} & \multicolumn{1}{r}{147} \\
Cc2 & \multicolumn{1}{r}{0.9083} & \multicolumn{1}{r}{1.00} & \multicolumn{1}{r}{1.00} & \multicolumn{1}{r}{0.8980} & \multicolumn{1}{r}{3.00} & \multicolumn{1}{r}{0.00} & \multicolumn{1}{r}{0.8974} & \multicolumn{1}{r}{1.15} & \multicolumn{1}{r}{0.00} & \multicolumn{1}{r}{0.8894} & \multicolumn{1}{r}{16.00} & \multicolumn{1}{r}{0.00} & \multicolumn{1}{r}{0.9083} & \multicolumn{1}{r}{164} \\
Cc3 & \multicolumn{1}{r}{0.9043} & \multicolumn{1}{r}{1.00} & \multicolumn{1}{r}{1.00} & \multicolumn{1}{r}{0.8972} & \multicolumn{1}{r}{1.30} & \multicolumn{1}{r}{0.00} & \multicolumn{1}{r}{0.8944} & \multicolumn{1}{r}{1.11} & \multicolumn{1}{r}{0.00} & \multicolumn{1}{r}{0.8923} & \multicolumn{1}{r}{1.29} & \multicolumn{1}{r}{0.00} & \multicolumn{1}{r}{0.9043} & \multicolumn{1}{r}{128} \\
Cc4 & \multicolumn{1}{r}{0.9087} & \multicolumn{1}{r}{1.00} & \multicolumn{1}{r}{1.00} & \multicolumn{1}{r}{0.9059} & \multicolumn{1}{r}{1.03} & \multicolumn{1}{r}{0.00} & \multicolumn{1}{r}{0.9046} & \multicolumn{1}{r}{1.05} & \multicolumn{1}{r}{0.00} & \multicolumn{1}{r}{0.9013} & \multicolumn{1}{r}{1.20} & \multicolumn{1}{r}{0.00} & \multicolumn{1}{r}{0.9087} & \multicolumn{1}{r}{14} \\
Cc5 & \multicolumn{1}{r}{0.9014} & \multicolumn{1}{r}{1.00} & \multicolumn{1}{r}{1.00} & \multicolumn{1}{r}{0.8996} & \multicolumn{1}{r}{1.02} & \multicolumn{1}{r}{0.00} & \multicolumn{1}{r}{0.8980} & \multicolumn{1}{r}{1.04} & \multicolumn{1}{r}{0.00} & \multicolumn{1}{r}{0.8972} & \multicolumn{1}{r}{1.05} & \multicolumn{1}{r}{0.00} & \multicolumn{1}{r}{0.9014} & \multicolumn{1}{r}{8} \\
Cc6 & \multicolumn{1}{r}{0.8980} & \multicolumn{1}{r}{1.00} & \multicolumn{1}{r}{1.00} & \multicolumn{1}{r}{0.8962} & \multicolumn{1}{r}{1.02} & \multicolumn{1}{r}{0.00} & \multicolumn{1}{r}{0.8949} & \multicolumn{1}{r}{1.03} & \multicolumn{1}{r}{0.00} & \multicolumn{1}{r}{0.8931} & \multicolumn{1}{r}{1.21} & \multicolumn{1}{r}{0.00} & \multicolumn{1}{r}{0.8980} & \multicolumn{1}{r}{12} \\
Cc7 & \multicolumn{1}{r}{0.9001} & \multicolumn{1}{r}{1.00} & \multicolumn{1}{r}{0.97} & \multicolumn{1}{r}{0.8940} & \multicolumn{1}{r}{1.09} & \multicolumn{1}{r}{0.00} & \multicolumn{1}{r}{0.8923} & \multicolumn{1}{r}{1.08} & \multicolumn{1}{r}{0.03} & \multicolumn{1}{r}{0.8916} & \multicolumn{1}{r}{1.23} & \multicolumn{1}{r}{0.00} & \multicolumn{1}{r}{0.9001} & \multicolumn{1}{r}{86} \\
Cc8 & \multicolumn{1}{r}{0.9042} & \multicolumn{1}{r}{1.00} & \multicolumn{1}{r}{1.00} & \multicolumn{1}{r}{0.8981} & \multicolumn{1}{r}{1.19} & \multicolumn{1}{r}{0.00} & \multicolumn{1}{r}{0.8955} & \multicolumn{1}{r}{1.09} & \multicolumn{1}{r}{0.00} & \multicolumn{1}{r}{0.8957} & \multicolumn{1}{r}{1.23} & \multicolumn{1}{r}{0.00} & \multicolumn{1}{r}{0.9042} & \multicolumn{1}{r}{65} \\
Cc9 & \multicolumn{1}{r}{0.9018} & \multicolumn{1}{r}{1.00} & \multicolumn{1}{r}{1.00} & \multicolumn{1}{r}{0.8932} & \multicolumn{1}{r}{1.11} & \multicolumn{1}{r}{0.00} & \multicolumn{1}{r}{0.8922} & \multicolumn{1}{r}{1.10} & \multicolumn{1}{r}{0.00} & \multicolumn{1}{r}{0.8905} & \multicolumn{1}{r}{1.22} & \multicolumn{1}{r}{0.00} & \multicolumn{1}{r}{0.9018} & \multicolumn{1}{r}{89} \\ \midrule
$Q_{25}$ & \multicolumn{1}{r}{0.9045} & \multicolumn{1}{r}{1.00} & \multicolumn{1}{r}{1.00} & \multicolumn{1}{r}{0.8996} & \multicolumn{1}{r}{1.11} & \multicolumn{1}{r}{0.00} & \multicolumn{1}{r}{0.8976} & \multicolumn{1}{r}{1.08} & \multicolumn{1}{r}{0.00} & \multicolumn{1}{r}{0.8925} & \multicolumn{1}{r}{1.29} & \multicolumn{1}{r}{0.00} & \multicolumn{1}{r}{0.9045} & \multicolumn{1}{r}{59} \\
$Q_{50}$ & \multicolumn{1}{r}{$\bm{ 0.9095}$} & \multicolumn{1}{r}{$\bm{1.00}$} & \multicolumn{1}{r}{$\bm{1.00}$} & \multicolumn{1}{r}{0.9055} & \multicolumn{1}{r}{1.18} & \multicolumn{1}{r}{0.00} & \multicolumn{1}{r}{0.9019} & \multicolumn{1}{r}{1.11} & \multicolumn{1}{r}{0.00} & \multicolumn{1}{r}{0.8985} & \multicolumn{1}{r}{1.35} & \multicolumn{1}{r}{0.00} & \multicolumn{1}{r}{$\bm{0.9095}$} & \multicolumn{1}{r}{79} \\
$Q_{75}$ & \multicolumn{1}{r}{0.9588} & \multicolumn{1}{r}{1.00} & \multicolumn{1}{r}{1.00} & \multicolumn{1}{r}{0.9557} & \multicolumn{1}{r}{1.32} & \multicolumn{1}{r}{0.00} & \multicolumn{1}{r}{0.9532} & \multicolumn{1}{r}{1.17} & \multicolumn{1}{r}{0.00} & \multicolumn{1}{r}{0.9506} & \multicolumn{1}{r}{1.96} & \multicolumn{1}{r}{0.00} & \multicolumn{1}{r}{0.9588} & \multicolumn{1}{r}{106} \\
$\mu$ & \multicolumn{1}{r}{0.9292} & \multicolumn{1}{r}{1.00} & \multicolumn{1}{r}{1.00} & \multicolumn{1}{r}{0.9251} & \multicolumn{1}{r}{1.32} & \multicolumn{1}{r}{0.00} & \multicolumn{1}{r}{0.9225} & \multicolumn{1}{r}{1.15} & \multicolumn{1}{r}{0.00} & \multicolumn{1}{r}{0.9178} & \multicolumn{1}{r}{4.35} & \multicolumn{1}{r}{0.00} & \multicolumn{1}{r}{0.9292} & \multicolumn{1}{r}{82} \\
$\sigma$ & \multicolumn{1}{r}{0.0303} & \multicolumn{1}{r}{0.00} & \multicolumn{1}{r}{0.01} & \multicolumn{1}{r}{0.0315} & \multicolumn{1}{r}{0.42} & \multicolumn{1}{r}{0.00} & \multicolumn{1}{r}{0.0320} & \multicolumn{1}{r}{0.17} & \multicolumn{1}{r}{0.01} & \multicolumn{1}{r}{0.0315} & \multicolumn{1}{r}{6.50} & \multicolumn{1}{r}{0.00} & \multicolumn{1}{r}{0.0303} & \multicolumn{1}{r}{41} \\ \bottomrule
\end{tabular}
\\[0.5em]
$Q_{50}$ denotes the median. The best median-values are bold. $Q_{25}$ and $Q_{75}$ denote the lower and upper quartile, respectively.

$\mu$ and $\sigma$ denote mean the standard deviation, respectively.
\end{table*}

\subsection{Runtime behavior}
\label{sec:test_runtime}
To compare the runtime of the three heuristics from the literature with the proposed heuristic PNS, a \emph{target} hypervolume value is defined for each instance. The runtime needed to achieve the target value is measured.

The target value is defined as the lowest $I_{HV}$ value per instance shown in Tab. \ref{tab:results_S} and Tab. \ref{tab:results_ABC}. Therefore, we are sure that each heuristic is able to reach the target value.
Using the best known hypervolume value as target value seems unsuitable because most of the tested heuristics are not able to attain this value constantly, which would limit the value of the experiment.
Following, each instance is solved by each heuristic 75 times and the time to target is measured. The total runtime per heuristic was limited to three minutes (180s). Note, if an algorithm could not reach the target value within 180s, then a runtime of 180s is reported anyway. Therefore, an algorithm might appear faster than it actually is. However, this behavior occurred only with the heuristic SPEA2A and never with the heuristic PNS.

The aggregated results are reported in Tab. \ref{tab:runtime}. According to the reported median values for the 37 instances, the new heuristic PNS ranks second. The fastest method is PGRASP$_P$+HPR, third place goes to PGRASP$_Q$+HPR, and fourth place goes to SPEA2A. The median runtime of the method PGRASP$_Q$+HPR for the larger instances is around 135s, which can be explained by a switch from the neighborhood search phase towards the path relinking phase, which is time dependent. Furthermore, although SPEA2A can repeatedly not achieve the target value in the predefined 180s (cf. $Q_{75}$ in Tab. \ref{tab:runtime}), for some of the larger instances SPEA2A seems competitive (cf. $Q_{25}$ and $Q_{50}$).

\begin{table*}[!htbp]
\caption{Comparison of the runtime (s) of the four heuristics for instance groups S, A, B, and C.}
\label{tab:runtime}
\scriptsize
\centering
\begin{tabular}{lllllllllllll} \toprule
Group & \multicolumn{3}{c}{PNS} & \multicolumn{3}{c}{PGRASP$_P$+HPR} & \multicolumn{3}{c}{PGRASP$_Q$+HPR} & \multicolumn{3}{c}{SPEA2A} \\
\cmidrule(lr){2-4}\cmidrule(lr){5-7}\cmidrule(lr){8-10}\cmidrule(lr){11-13}
 & \multicolumn{1}{c}{$Q_{25}$} & \multicolumn{1}{c}{$Q_{50}$} & \multicolumn{1}{c}{$Q_{75}$} & \multicolumn{1}{c}{$Q_{25}$} & \multicolumn{1}{c}{$Q_{50}$} & \multicolumn{1}{c}{$Q_{75}$} & \multicolumn{1}{c}{$Q_{25}$} & \multicolumn{1}{c}{$Q_{50}$} & \multicolumn{1}{c}{$Q_{75}$} & \multicolumn{1}{c}{$Q_{25}$} & \multicolumn{1}{c}{$Q_{50}$} & \multicolumn{1}{c}{$Q_{75}$} \\ \midrule
S & \multicolumn{1}{r}{0.03} & \multicolumn{1}{r}{0.07} & \multicolumn{1}{r}{0.12} & \multicolumn{1}{r}{0.02} & \multicolumn{1}{r}{$\bm{0.06}$} & \multicolumn{1}{r}{0.2} & \multicolumn{1}{r}{0.06} & \multicolumn{1}{r}{0.16} & \multicolumn{1}{r}{0.635} & \multicolumn{1}{r}{104.16} & \multicolumn{1}{r}{168.36} & \multicolumn{1}{r}{182.67} \\
A & \multicolumn{1}{r}{2.64} & \multicolumn{1}{r}{3.66} & \multicolumn{1}{r}{5.04} & \multicolumn{1}{r}{0.30} & \multicolumn{1}{r}{$\bm{0.58}$} & \multicolumn{1}{r}{1.82} & \multicolumn{1}{r}{88.36} & \multicolumn{1}{r}{135.05} & \multicolumn{1}{r}{145.61} & \multicolumn{1}{r}{183.05} & \multicolumn{1}{r}{182.94} & \multicolumn{1}{r}{183.20} \\
B & \multicolumn{1}{r}{7.86} & \multicolumn{1}{r}{11.75} & \multicolumn{1}{r}{18.48} & \multicolumn{1}{r}{0.91} & \multicolumn{1}{r}{$\bm{1.80}$} & \multicolumn{1}{r}{3.89} & \multicolumn{1}{r}{10.81} & \multicolumn{1}{r}{135.05} & \multicolumn{1}{r}{135.19} & \multicolumn{1}{r}{8.14} & \multicolumn{1}{r}{97.55} & \multicolumn{1}{r}{182.95} \\
C & \multicolumn{1}{r}{24.07} & \multicolumn{1}{r}{35.28} & \multicolumn{1}{r}{52.54} & \multicolumn{1}{r}{3.50} & \multicolumn{1}{r}{$\bm{10.20}$} & \multicolumn{1}{r}{36.67} & \multicolumn{1}{r}{3.74} & \multicolumn{1}{r}{135.05} & \multicolumn{1}{r}{137.79} & \multicolumn{1}{r}{13.61} & \multicolumn{1}{r}{26.27} & \multicolumn{1}{r}{182.73} \\
S. A. B. C & \multicolumn{1}{r}{3.50} & \multicolumn{1}{r}{16.65} & \multicolumn{1}{r}{34.37} & \multicolumn{1}{r}{0.52} & \multicolumn{1}{r}{$\bm{2.39}$} & \multicolumn{1}{r}{10.75} & \multicolumn{1}{r}{1.45} & \multicolumn{1}{r}{87.66} & \multicolumn{1}{r}{135.17} & \multicolumn{1}{r}{14.30} & \multicolumn{1}{r}{142.33} & \multicolumn{1}{r}{182.87} \\ \bottomrule
\end{tabular}%
\end{table*}

In the appendix, we discuss empirical runtime distributions based on six selected instances (cf. Fig.~\ref{fig:pns-runtime}a -- Fig.~\ref{fig:pns-runtime}f) in order to give more insights into the runtime behavior of the heuristics.

\section{Conclusion}
\label{sec:conclusion}

Considering quality aspects during winner determination in a combinatorial reverse auction for transport contracts is of practical importance. In this paper, we  studied a bi-objective winner determination problem that is based on the set covering problem and minimizes the total transport costs and the total transport quality simultaneously. To solve this problem, the heuristic PNS was developed. PNS is inspired by the metaheuristics GRASP and ALNS.
To construct an initial set of non dominated solutions, PNS applies a dominance-based randomized greedy heuristic which uses a \emph{two-stage candidate bid selection procedure}.
This idea to greedily construct an initial set of non-dominated solution is effective and may be well suited to solve other multi objective combinatorial optimization problems.
The set of constructed solutions is improved by means of a search in large neighborhoods which switches the applied parameters (removal probability of bids and greedy rating function) in a self-adaptive manner. Self-adaptive configurations depend on individual solutions and not on the entire approximation set. This is an important feature because the structure of solutions from the approximated Pareto front may be very different and this strongly influences the choice of a suitable destroy or repair operator.
 PNS was tested by means of 37 benchmark instances. In terms of approximation set quality, PNS outperforms all known heuristics on each of the 37 benchmark instances. Furthermore, PNS is the second fastest method tested.
Subject of our future research will be the development of solution approaches for bi-objective winner determination problems which take into account additional business constraints proposed e.g. by \citet{Caplice_2006}.

%% The Appendices part is started with the command \appendix;
%% appendix sections are then done as normal sections
\appendix
\section{Time to target plots}
\label{sec:appendix}

The runtime of stochastic algorithms can be compared by empirical runtime distributions (cf. \citet{Hoos_1999}, \citet{Ribeiro_2009}).
The time required by a stochastic algorithm to find a solution that achieves a specified minimum quality (\emph{target value}) is interpreted as a stochastic variable denoted \emph{time to target}. In the following, we visualise some empirical runtime distributions by means of \emph{time to target} plots which have been introduced by \citet{Feo_1994}. To draw the plots the programme of \citet{Aiex_2007} was used.

\begin{figure}[!htbp]%
\centering
\subfloat[Instance Aa2, target value $I_{HV} = 0.9131$]{
\label{fig:nbs-runtime-Aa2}
\includegraphics[scale=0.35]{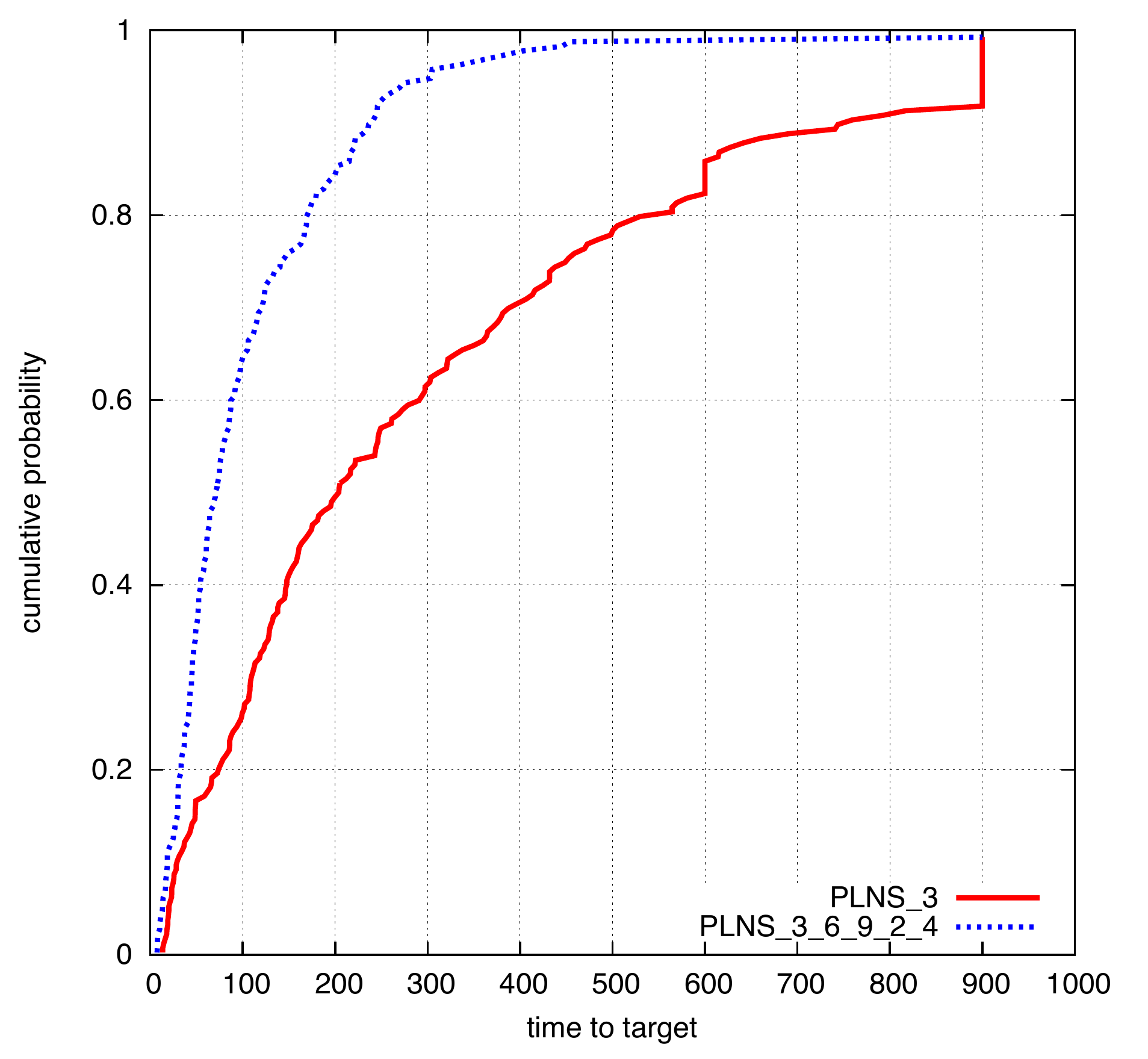}
}\qquad
\subfloat[Instance Bb6, target value $I_{HV} = 0.9097$]{
\label{fig:nbs-runtime-Bb6}
\includegraphics[scale=0.35]{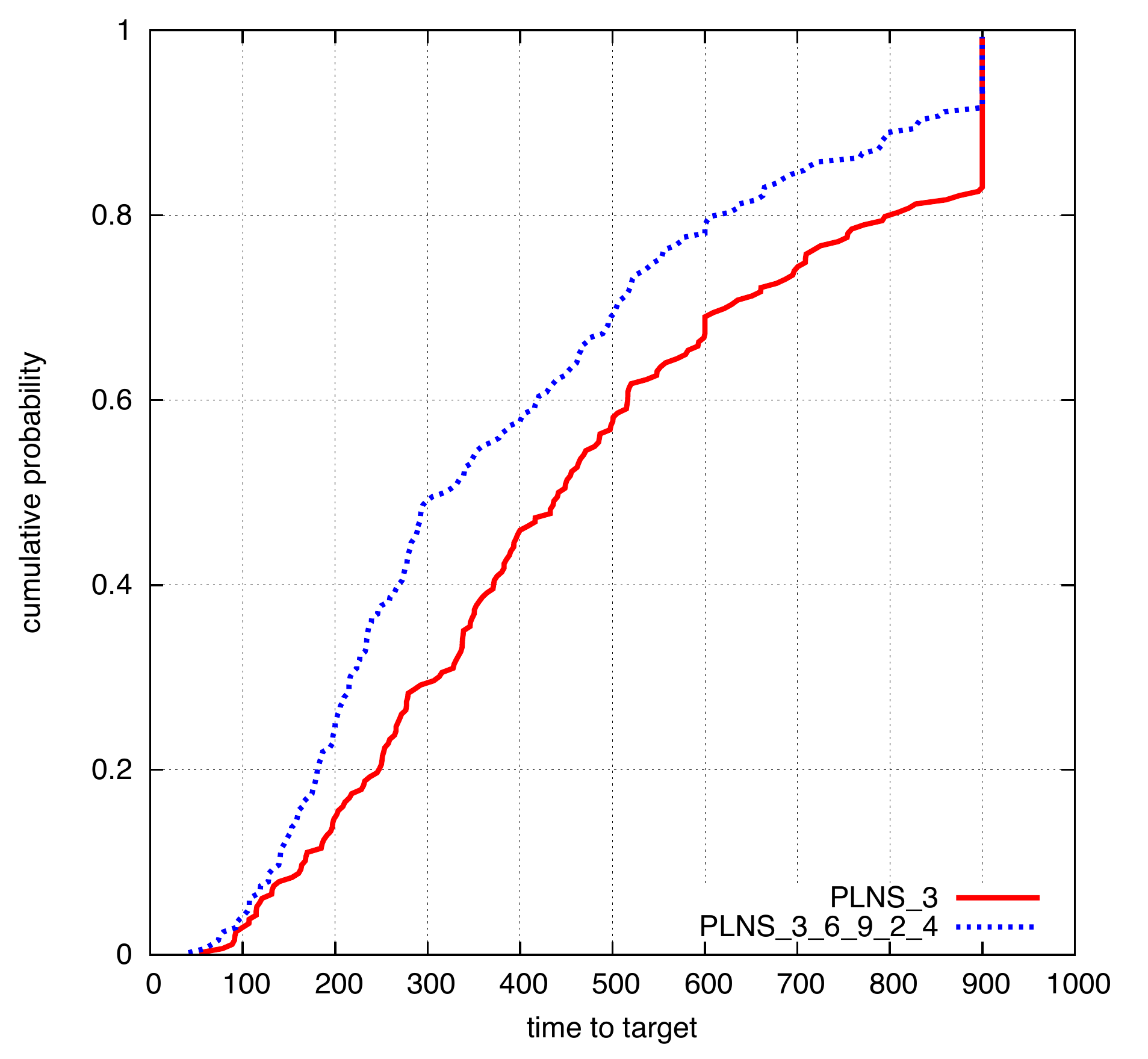}
}\\
\subfloat[Instance Cc5, target value $I_{HV} = 0.9046$]{
\label{fig:nbs-runtime-Cc5}
\includegraphics[scale=0.35]{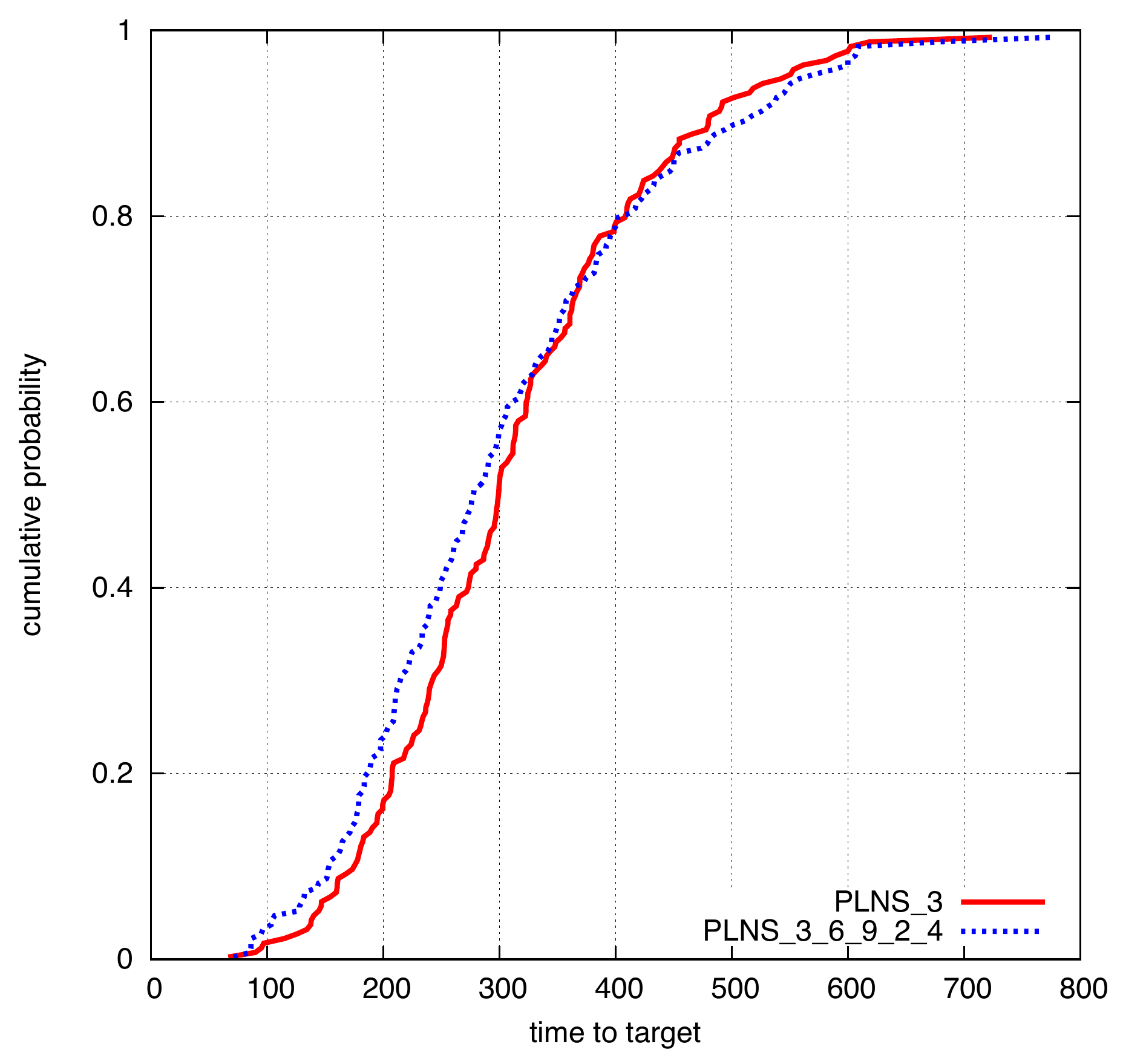}}%
\caption{Empirical runtime distribution of PLNS$_{\bm d = (3)}$ and PLNS$_{\bm d = (3,6,9,2,4)}$ (dotted) determined by 200 runs.}
\label{fig:nbs-runtime}
\end{figure}

The evaluation of destroy strategies of Sect.~\ref{sec:res_improvement} is supplemented by the time to target plots of Fig.~\ref{fig:nbs-runtime-Aa2} to Fig.~\ref{fig:nbs-runtime-Cc5}. We compare the destroy strategies $\bm d = (3)$ and $\bm d = (3, 6, 9, 2, 4)$  used in the improvement heuristic PLNS.
In the plot of Fig.~\ref{fig:nbs-runtime-Aa2}, for example, we observer that the strategy $(3, 6, 9, 2, 4)$ achieves the target value in less than 300 seconds with a probability of about 95 percent, while the strategy $(3)$ accomplishes a probability of only about 62 percent to reach the same target value within the same time. The advantage of $(3, 6, 9, 2, 4)$ over $(3)$ persists but decreases for larger instances (cf. Fig.~\ref{fig:nbs-runtime-Bb6}, Fig.~\ref{fig:nbs-runtime-Cc5}).

\begin{figure*}[!htbp]%
\centering
\subfloat[Instance Aa2]{
\label{fig:runtime-Aa2}
\includegraphics[scale=0.35]{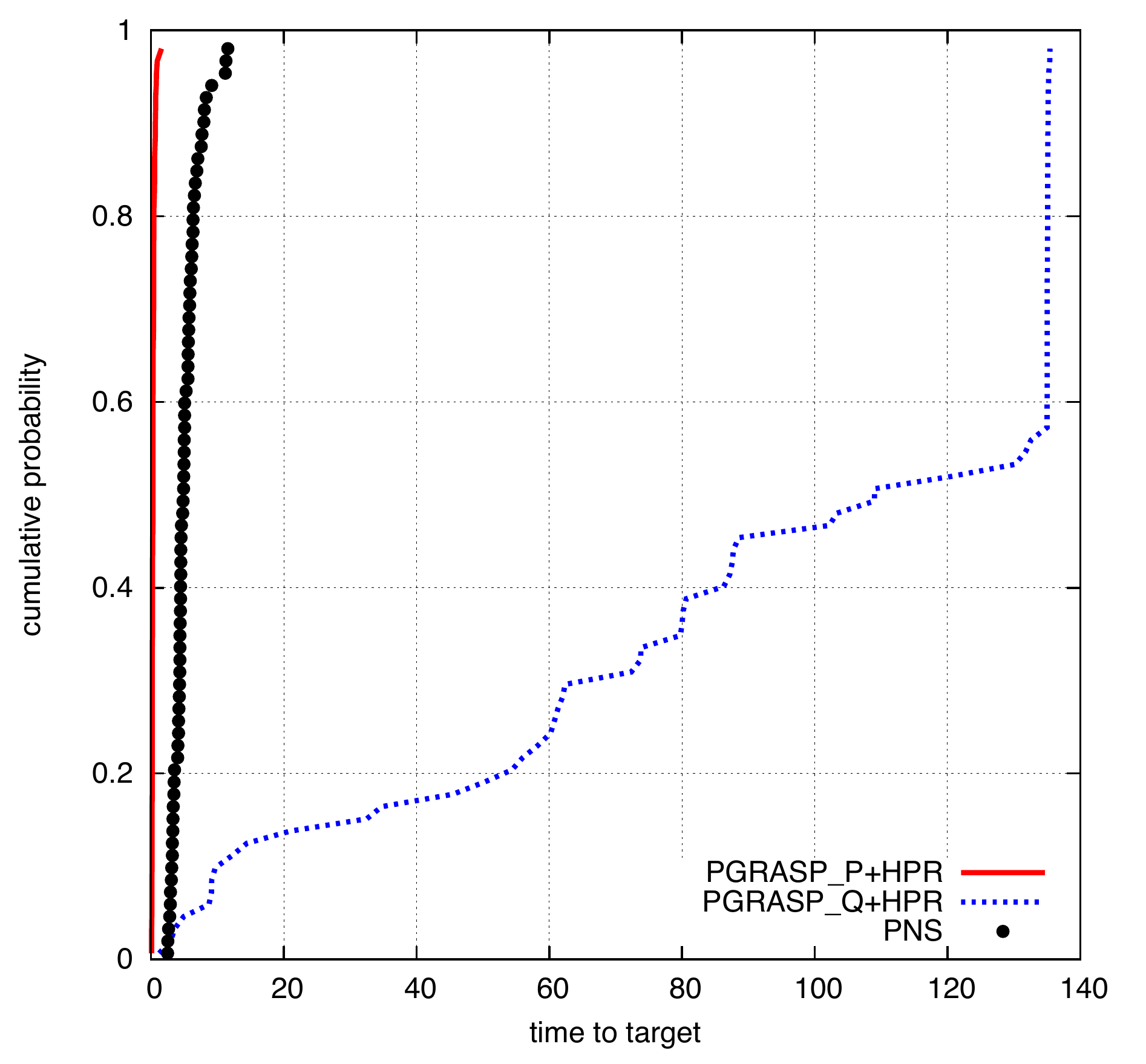}
}\quad
\subfloat[Instance Bb2]{
\label{fig:runtime-Bb2}
\includegraphics[scale=0.35]{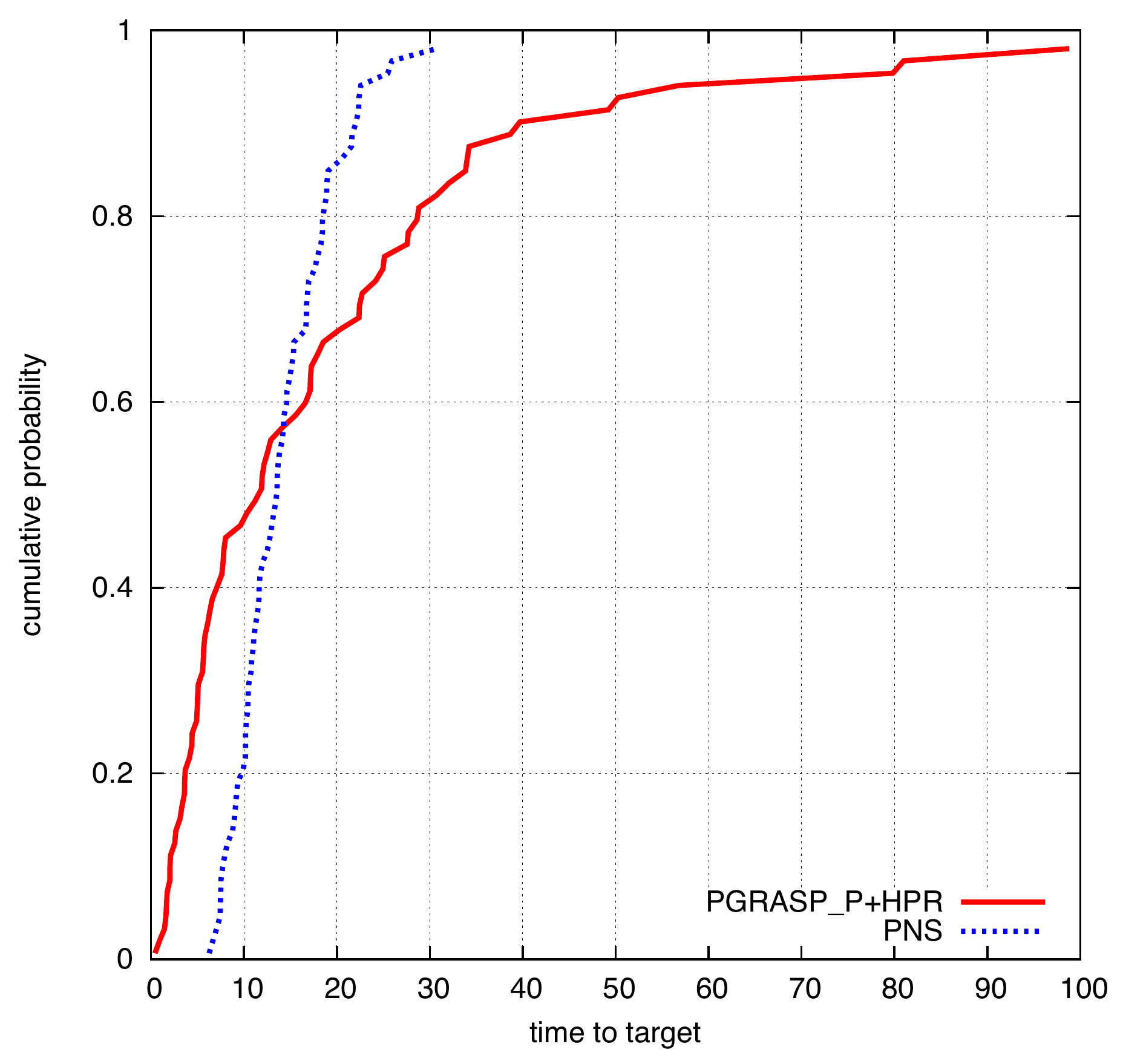}
}\quad
\subfloat[Instance Bb4]{
\label{fig:runtime-Bb4}
\includegraphics[scale=0.35]{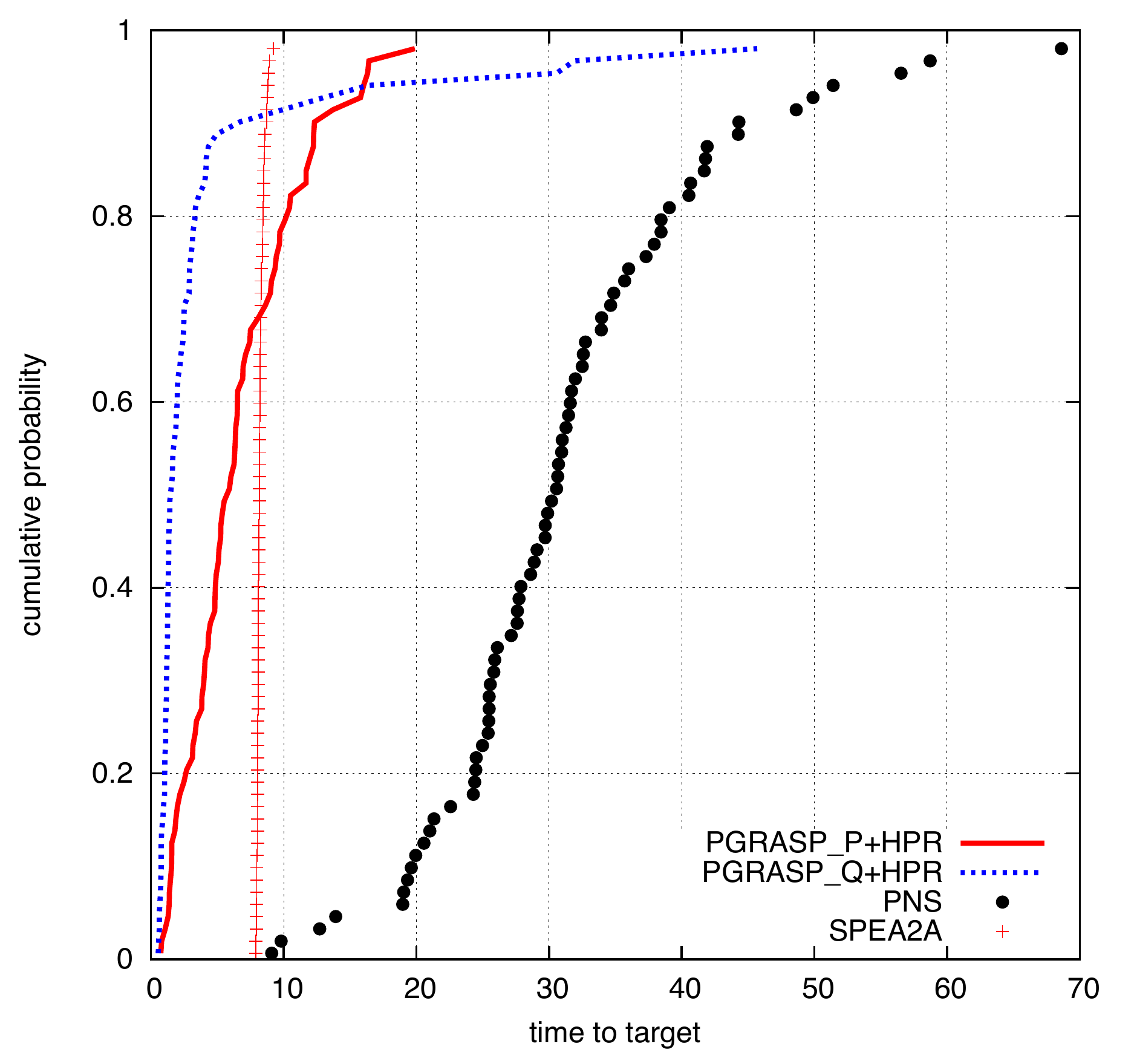}
}\quad
\subfloat[Instance Ca3]{
\label{fig:runtime-Ca3}
\includegraphics[scale=0.35]{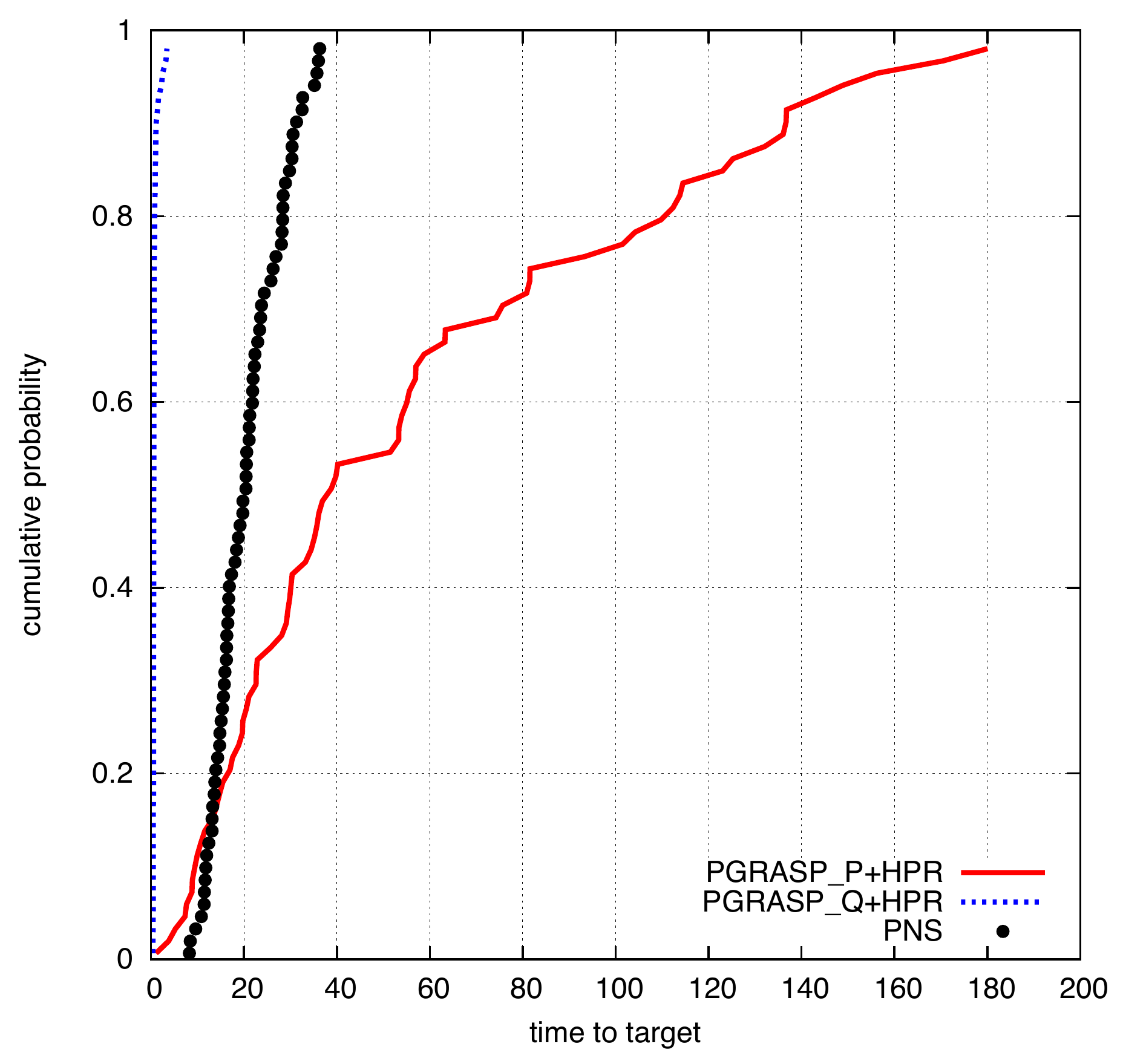}
}\quad
\subfloat[Instance Cc6]{
\label{fig:runtime-Cc6}
\includegraphics[scale=0.35]{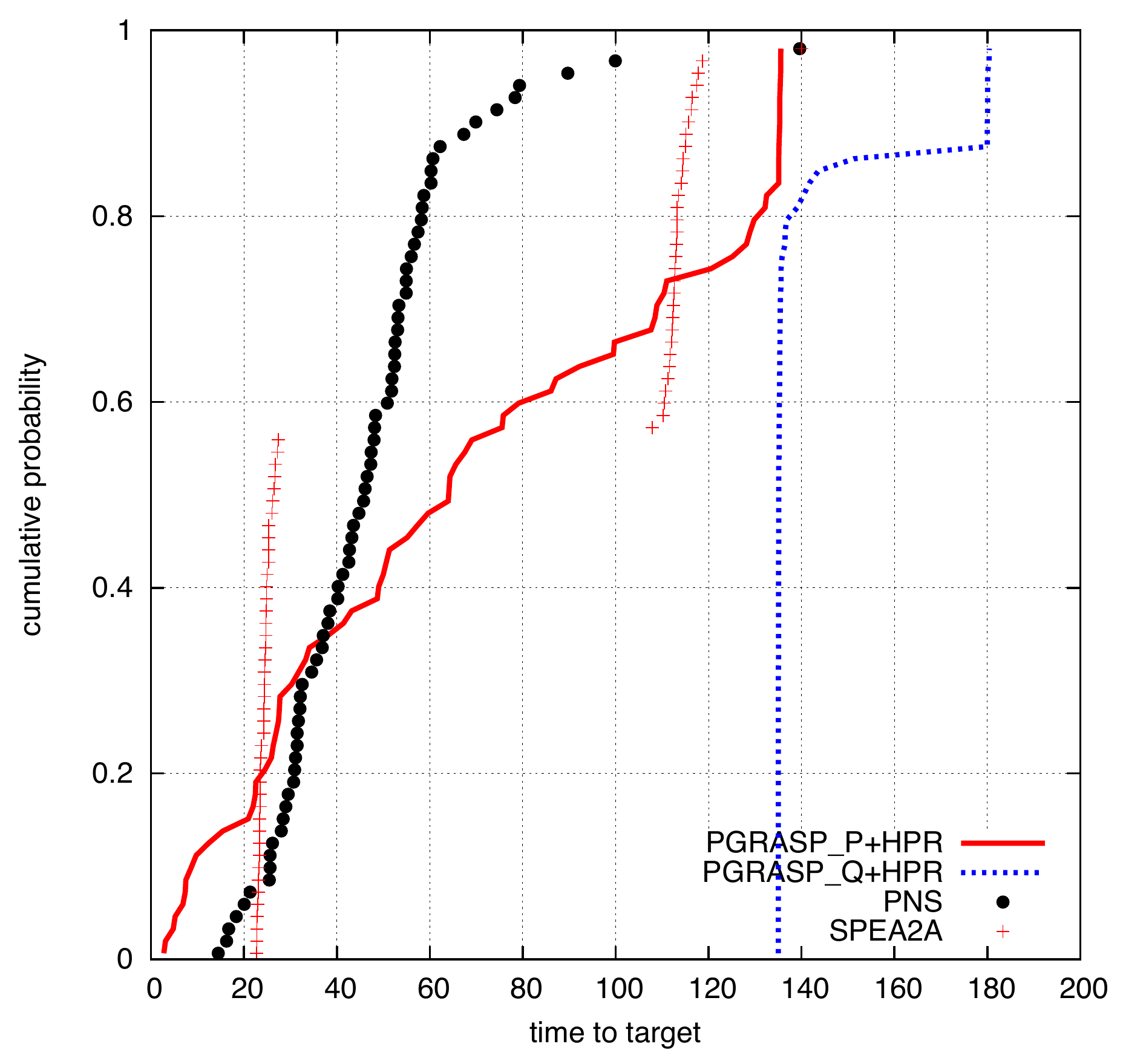}
}\quad
\subfloat[Instance Cc9]{
\label{fig:runtime-Cc9}
\includegraphics[scale=0.35]{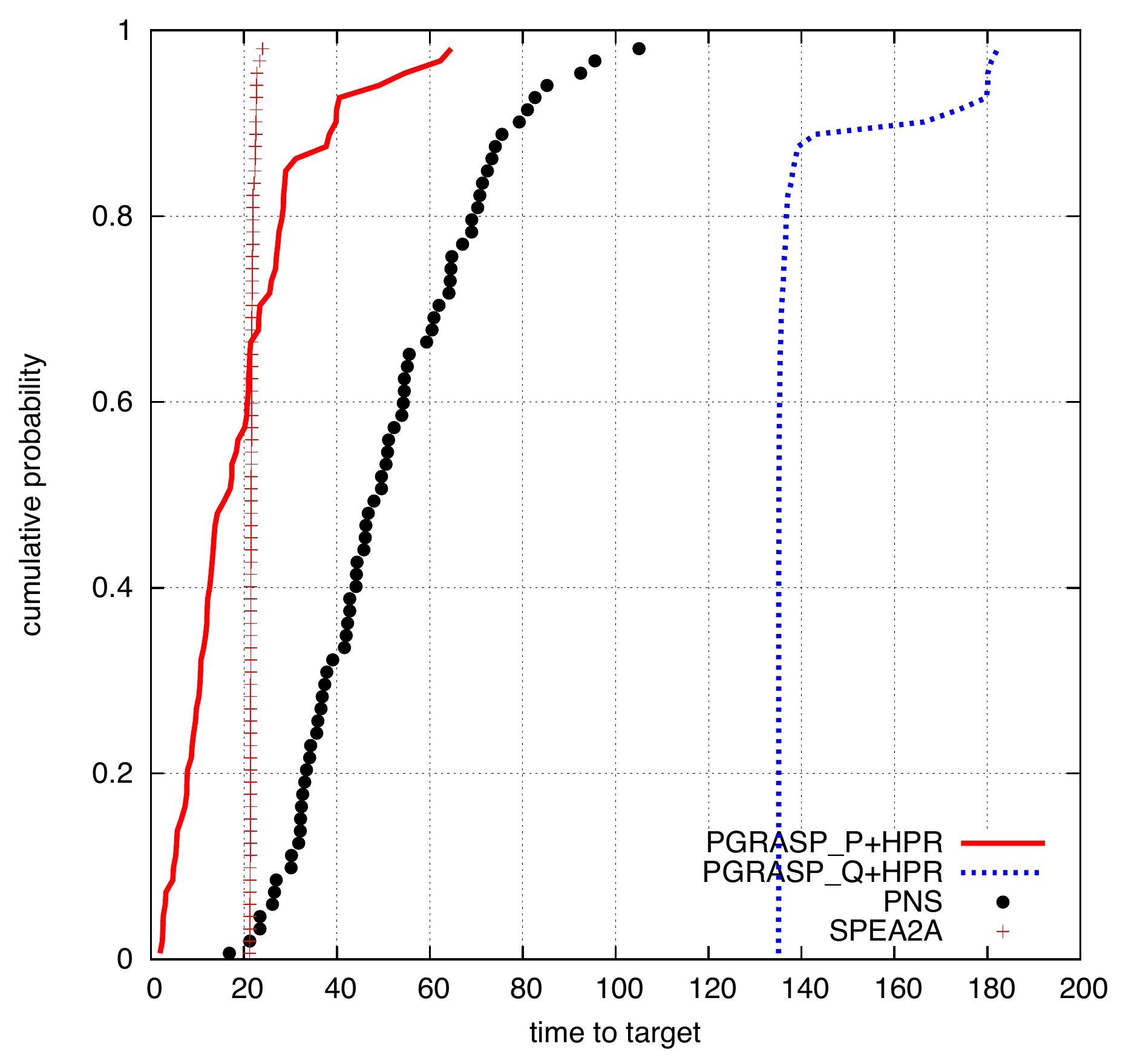}
}
\caption{Empirical runtime distribution of PNS compared to benchmark heuristics determined by 75 runs}
\label{fig:pns-runtime}
\end{figure*}

All in all, PNS is the second fastest heuristic tested after PGRASP$_P$+HPR (cf. Sect.~\ref{sec:test_comparison}).
This conclusion is also supported by the plots of Fig.~\ref{fig:runtime-Aa2} and Fig.~\ref{fig:runtime-Bb2}. However, there are other plots like Fig.~\ref{fig:runtime-Bb4} -- Fig.~\ref{fig:runtime-Cc9}. These are not representative for the 37 benchmark instances tested but they were chosen in order to contribute to a differentiated assessment of runtime performance. For example, SPEA2A is usually far out with respect to both solution quality and runtime performance. Therefore in some of the time to target plots the respective curve is missing. However, for the instances Cc6 and Cc9 SPEA2A achieves quite good results. With respect to runtime performance the heuristic PGRASP$_Q$+HPR ranks third but in some cases it significantly outperforms PGRASP$_P$+HPR (cf. Fig.~\ref{fig:runtime-Ca3}). Nonetheless, one should always keep in mind that time to target plots depend on the chosen target value which were set as described in Sect.~\ref{sec:test_runtime}.

%% References
%%
%% Following citation commands can be used in the body text:
%% Usage of \cite is as follows:
%%   \cite{key}          ==>>  [#]
%%   \cite[chap. 2]{key} ==>>  [#, chap. 2]
%%   \citet{key}         ==>>  Author [#]

%% References with bibTeX database:

\bibliographystyle{model3-num-names}
%\bibliography{../../../Latex/texmf/bibtex/bib/literaturdatenbank}

\end{document}